\documentstyle[aps,eqsecnum,preprint]{revtex}
\tightenlines
\begin{document}

\title{Spin dynamics simulation of the three-dimensional XY model:
Structure factor and transport properties}
\author{M. Krech$^{\dag}$ and D.P. Landau$^{\ddag}$ \\
$^{\dag}$Institut f\"ur Theoretische Physik, RWTH Aachen, 52056 Aachen,
Germany \\
$^{\ddag}$Center for Simulational Physics, University of Georgia, Athens,
Georgia 30602, USA}
\maketitle

\begin{abstract}
We present extensive Monte-Carlo spin dynamics simulations of the classical
XY model in three dimensions on a simple cubic lattice with periodic boundary
conditions. A recently developed efficient integration algorithm for the
equations of motion is used, which allows a substantial improvement of
statistics and large integration times. We find spin wave peaks in a wide
range around the critical point and spin diffusion for all temperatures. At
the critical point we find evidence for a violation of dynamic scaling in
the sense that independent components of the dynamic structure factor
$S({\bf q},\omega)$ require different dynamic exponents in order to obtain
scaling. Below the critical point we investigate the dispersion relation of
the spin waves and the linewidths of $S({\bf q},\omega)$ and find agreement
with mode coupling theory. Apart from strong spin wave peaks we observe
additional peaks in $S({\bf q},\omega)$ which can be attributed to two-spin
wave interactions. The overall lineshapes are also discussed and compared
to mode coupling predictions. Finally, we present first results for the
transport coefficient $D(q,\omega)$ of the out-of-plane magnetization
component at the critical point, which is related to the thermal conductivity
of $^4$He near the superfluid-normal transition.
\noindent
\end{abstract}
\draft


\section{Introduction}
The theoretical investigation of classical spin systems has played a key role
in the understanding of phase transitions, critical behavior, scaling, and
universality \cite{Amit78,Parisi88}. In particular, the claasical Ising, the
XY, and the Heisenberg model are the most relevant spin models in three
dimensions. Each of these simple models represents a universality class which,
apart from the spatial dimensionality and the range of the interactions,
is characterized by the number of components $N$ of the order parameter,
e.g, the magnetization in the case of ferromagnetic models.
Despite their simplicity these spin systems continue to be of high relevance
within the framework of {\em dynamic} behavior near critical points
\cite{HH77} (see also Ref.\cite{DPLMK98} for a recent review). The Ising
$(N = 1)$, the XY $(N = 2)$, and the Heisenberg $(N = 3)$ universality
class can be extended towards {\em dynamic} universality classes, which
in addition to their static properties are characterized by the set of
conservation laws \cite{HH77}. Special attention must be paid to the
presence of energy conserving driving terms in the
equations of motion which lead to propagating modes (spin waves) below the
critical temperature and thus modify the dynamics \cite{HH77}. The discrete
nature of Ising spins does not allow such terms so that
its dynamics is always of relaxational type (see Ref.\cite{HH77} for a
complete classification).

The simplest spin model which allows propagating modes is a particular
version of the ferromagnetic XY model $(N = 2)$. The dynamics here is
characterized by a nonconserved order parameter which is  dynamically
coupled to a {\em conserved} quantity (see Sec.II). The presence
of spin waves reduces the value of the dynamic critical exponent $z$ as
compared to pure critical relaxation \cite{HH77}. The same is true for the
isotropic Heisenberg model for which the $N = 3$ component magnetization
vector is always conserved in the presence of energy conserving driving
terms. If the model is ferromagnetic, the magnetization is the order
parameter. However, for an {\em antiferromagnet}, the
magnetization plays the role of a conserved vector which is dynamically
coupled to the {\em nonconserved} order parameter (staggered magnetization).
This difference in the conservation laws causes the classical Heisenberg
ferro- and antiferromagnet to be in different dynamic universality classes
although they belong to the same static universality class.
Due to their fundamental role in the understanding of the critical dynamics
in magnets Heisenberg ferro- and antiferromagnets have been thoroughly
studied analytically by mode coupling theories (see Ref.\cite{Kawa76} for a
general overview) especially in presence of dipolar interactions
\cite{FS94} and numerically by spin dynamics in $d = 2$ \cite{Keren94} and
in $d = 3$ \cite{CL94,BCL96} and by methods closely related to molecular
dynamics \cite{RapLan96}.

The XY model may be viewed as a Heisenberg ferromagnet with an easy-plane
$(xy)$ anisotropy such that the order parameter has only two components.
Planar ferromagnets are realized by layered compounds such as K$_2$CuF$_4$
\cite{HY79} and Rb$_2$CrCl$_4$ \cite{HDJP86} which almost act as
two-dimensional systems. The best results available today have been obtained
on CoCl$_2$ intercalated in graphite \cite{WZS94}, where a crossover from
two-dimensional to three-dimensional behavior in the correlations has been
observed below the Kosterlitz-Thouless temperature.
Apart from the evident interpretation as a planar
ferromagnet the XY model captures a larger variety of phenomena than the
Ising or the Heisenberg model.
Despite its continuous $O(2)$ symmetry the XY model undergoes
a continuous phase transition at a finite temperature in two dimensions,
known as the Kosterlitz - Thouless transition
\cite{KT72}. Rather than by the onset of long-ranged order the transition is
solely characterized by a diverging correlation length, when the critical
temperature is approached from above. Due to a peculiar conspiracy between
the spatial dimensionality $d = 2$ and the number of spin components
$N = 2$ configurations of bound and free vortices dominate the critical
behavior of the XY model, where the unbinding of vortex pairs marks the
point of the phase transition \cite{KT72}. Naturally, many attempts to
describe the critical dynamics of the XY model in $d = 2$ theoretically
are based on the dynamics of vortices and vortex pairs
\cite{Huber82,Mertens,Gouvea89,Volkel91}. According to analytical and
numerical investigations for the ferromagnetic case
\cite{Mertens,Gouvea89,Volkel91,DPLRG92} the in-plane component
$S_{xx}({\bf q},\omega)$ and the out-of-plane component
$S_{zz}({\bf q},\omega)$ of the dynamic structure factor are expected to
have central peaks above the transition. The line
shapes of these peaks are predicted to be squared Lorentzian and Gaussian,
respectively. Below the transition only spin-wave
peaks are expected \cite{Mertens}. To test these
specific predictions much numerical effort has been spent on spin
dynamics simulations of the XY model in $d = 2$ \cite{EvLan96,Costa}.
Although dynamic finite-size scaling and the value of the dynamical
exponent $z = 1$ has been confirmed to a high degree of confidence
\cite{EvLan96}, the measured lineshapes of $S({\bf q},\omega)$
\cite{EvLan96,Costa} are apparently not well captured by analytical
theories (see Ref.\cite{DPLMK98} for details). It is therefore instructive
to measure $S({\bf q},\omega)$ for the XY model in $d = 3$ for which
configurations of bound or free vortices do not play any special role for
the critical behavior and should therefore
not provide particularly noticable contributions to the structure factor.
In $d = 3$ the dynamics of the planar ferromagnet has been investigated by
mode coupling theory \cite{TFS91} and specific predictions concerning
line shapes and -widths have been made which can be compared with our
data (see Sec.IV).

It is well known that the $\lambda$-transition of $^4$He is the in the
XY universality class, but the applications of the XY model for the
physics of $^4$He reach far beyond that. The spin dynamics for the XY
model is the lattice analogue of the dynamical model E (symmetric
planar ferromagnet \cite{HH77}) which {\em asymptotically} also describes
the {\em critical dynamics} of $^4$He near the $\lambda$-line
\cite{HH77,HHS76,DP78,Dohm78}. If one therefore studies the transport
properties of the XY model near the critical point $T_c$, one should obtain
lattice analogues of the corresponding transport coefficients of $^4$He
near the $\lambda$-transition. In this respect the aforementioned conserved
quantity plays a particularly interesting part, because it is related to
the entropy density in $^4$He and its associated transport coefficient
corresponds to the thermal conductivity of $^4$He \cite{HHS76,Dohm78}
which is an experimentally accessible quantity \cite{Ahlers68}. Below
the critical temperature spin waves in the XY model then correspond to
travelling waves of second sound in $^4$He. These propagating modes cause
the thermal conductivity to diverge at the lambda transition of bulk
$^4$He \cite{HHS76,Dohm78}. In a finite system like our simulation sample,
one therefore expects critical finite-size rounding of the thermal
conductivity, which can be studied in the framework of the spin dynamics
simulation and which should also be observable in experiments.

To what extent the spin dynamics simulation actually captures the critical
dynamics of $^4$He is a rather delicate question. Although the {\em
asymptotic} behavior is described by model E, the actual {\em crossover}
to the asymptotic behavior, i.e., the decay of {\em nonasymptotic}
corrections is governed by the specific heat exponent
$\alpha \simeq -0.013$ \cite{LSNCI96}, which is so small that the true
asymptotic behavior will never be seen in a simulation. From the point of
view of analytic theory this means that in order to capture the
nonansymptotic effects present in experiments one has to replace model E
by the more complicated model F \cite{HH77,HHS76,Dohm78,Dohm91}. From the
point of view of spin dynamics simulations this means that one has to look
for sources of such nonasymptotic behavior artificially generated by the
simulation method and other nonasymptotic corrections not captured by the
model or the method (see Sec.II). Apart from these problems, it should also
be mentioned that the dynamical model E has two renormalization group fixed
points. One of these fixed points yields dynamic scaling with a single
dynamic exponent $z = d/2$ in $d$ dimensions, whereas the other gives rise
to a weak {\em violation} of dynamic scaling. Theoretical arguments
\cite{DP78,Dohm78,Dohm91}and experimental evidence indicate that
the latter fixed point is the stable one for $^4$He in $d = 3$, i.e.,
the critical dynamics is characterized by two different
dynamic exponents $z_{\phi}$ (order parameter) and $z_m$ (conserved
quantity) which fulfill the scaling relation $z_{\phi} + z_m = d$. Their
difference $\omega_w \equiv z_{\phi} - z_m \neq 0$ has the nature of a
dynamic Wegner exponent and is known as the {\em transient} exponent
\cite{Dohm78,Dohm91}.

The remainder of the paper is organized as follows. In Sec.II we present
the model and the simulation methods used to generate equilibrium
configurations and to obtain the critical point of the model and its
static critical exponents. Furthermore, the equations of motion and the
method used to integrate them numerically are presented. In Sec.III we
briefly discuss the static critical behavior of our model and present
an accurate estimate of the critical temperature. Sec.IV is devoted to
the discussion of the dynamic structure factor and the comparison with
predictions of analytic theory. In Sec.V we present first results for
the lattice analogue of the thermal conductivity and discuss its scaling
properties. A summary and prospects for future work are given in Sec.VI.
Unless otherwise stated statistical errors quoted in this work correspond
to one standard deviation.

\section{Model and simulation method}
The system under investigation is given by a ferromagnetic Heisenberg
model with the strongest possible easy-plane anisotropy. The model
Hamiltonian reads
\begin{equation}
\label{HXY}
{\cal H} = -J \sum_{\langle i j \rangle}
\left( S_i^x S_j^x + S_i^y S_j^y \right) \quad ,
\end{equation}
where $\langle i j \rangle$ denotes a nearest neighbor pair of spins on
a simple cubic lattice in three dimensions. The lattice contains $L$
lattice sites in each direction and in order to avoid surface effects
periodic boundary conditions are applied. Each spin ${\bf S}_i$ is a
classical spin ${\bf S}_i = \left(S_i^x,S_i^y,S_i^z\right)$
with the normalization $|{\bf S}_i| = 1$.
The easy-plane anisotropy in Eq.(\ref{HXY}) is the strongest possible
in the sense that the $z$-components of the spins do not
couple, so that Eq.(\ref{HXY}) looks like the standard Hamiltonian
for the usual (plane rotator) XY model.

As a starting point for the spin dynamics a sequence of equilibrium
configurations is needed to provide initial conditions for the equations
of motion. These configurations are obtained from a Monte-Carlo simulation
of the model Hamiltonian given by Eq.(\ref{HXY}). The Monte-Carlo algorithm
chosen is a hybrid scheme, where each hybrid Monte-Carlo step (MCS) consists
of 10 updates each of which can be one of: one Metropolis
sweep of the whole lattice, one single cluster Wolff update \cite{Wolff89},
or one overrelaxation update of the whole lattice \cite{CL94}. The Metropolis 
algorithm updates the lattice sequentially in the standard way.
According to the detailed balance condition we choose the acceptance
probability $p(\beta \Delta E) = 1 / (\exp(\beta \Delta E) + 1)$ for a single
spin flip, where $\Delta E$ is the change in configurational energy according
to Eq.(\ref{HXY}) and $\beta = 1/(k_B T)$.

The Wolff algorithm also works the standard way \cite{Wolff89}, except that
{\em only} the $x$ and $y$ components of the spins are used for the cluster
growth. This means that a cluster update never changes the
$z$-component of any spin so that the Wolff algorithm is nonergodic in
this case. Our cluster update is still a valid Monte-Carlo step in the
sense that it fulfills detailed balance, however, in order to provide a
valid Monte-Carlo algorithm it has to be used together with the Metropolis
algorithm described above in a hybrid fashion. The use of Wolff
updates allows us to take advantage of improved estimators \cite{Has90}
for magnetic quantities.

The overrelaxation part of
the algorithm performs a microcanonical update of the configuration in
the following way. The local configurational energy has the functional
form of a scalar product of the spins, where according to Eq.(\ref{HXY})
only the $x$ and $y$ components are involved. With respect to the sum
of its nearest neighbor spins each spin has a transverse component in the
$xy$ plane which does not enter the scalar product. The overrelaxation
algorithm simply scans the lattice sequentially, determines this
transverse component for each lattice site and flips its sign. This does
not change the local configurational energy $(\Delta E = 0)$ and by virtue
of the usual Metropolis acceptance function $f(\beta \Delta E) =
\min(\exp(-\beta \Delta E),1)$ the update is always accepted. Along with
this simple operation the sign of $S_i^z$ is flipped with probability $1/2$
at each lattice site which according to Eq.(\ref{HXY}) also does not change
the energy of the configuration. This overrelaxation algorithm is similar to
the one used in Ref.\cite{CL94} and it quite efficiently decorrelates
subsequent configurations over a wider range of temperatures around the
critical point than does the Wolff algorithm. Typically, we use three
Metropolis {\em (M)}, five single cluster Wolff {\em (C)}, and two
overrelaxation updates {\em (O)} in a hybrid Monte-Carlo step in the critical
region of our XY model. The inividual updates are mixed automatically in the
program so that the update sequence {\em (M C C M O C M C C O)} is generated
as one hybrid Monte-Carlo step in this case. The random number generator we
use is the shift register generator R1279 given by the recursion relation
$X_n = X_{n-p} \oplus X_{n-q}$ for $(p,q) = (1279,1063)$. Generators like
this are known to cause systematic errors in combination with the Wolff
algorithm \cite{cluerr}; however, for lags $(p,q)$ as large as the ones
used here these errors will be far smaller than typical statistical errors.
They are further reduced by the hybrid nature of our algorithm \cite{AMFDPL}.

The spin dynamics of the XY model is defined by the equations of motion
\begin{equation}
\label{XYeqmot}
{d \over dt}{\bf S}_k = {\partial {\cal H} \over \partial {\bf S}_k}
\times {\bf S}_k ,
\end{equation}
where ${\cal H}$ ist the Hamiltonian defined by Eq.(\ref{HXY}). One may
interpret Eq.(\ref{XYeqmot}) as the direct classical analogue of the
Heisenberg equations of motion for spin operators, where $\hbar = 1$ so
that energies and frequencies are measured in the same units. From the
symmetry of Eq.(\ref{HXY}) it is evident that the components $M_x$ and $M_y$
of the magnetization ${\bf M} = \sum_k {\bf S}_k$ are not conserved under
the dynamics given by Eq.(\ref{XYeqmot}). Note that the two-component vector
$(M_x,M_y)$ is the {\em order parameter} of the XY model. The $z$- or
{\em out-of-plane} component $M_z$ of the magnetization ${\bf M}$ is
just the conserved quantity within the framework of model E dynamics we
have already referred to in Sec.I \cite{HH77}. Note that Eq.(\ref{HXY}) is
invariant with respect to the transformation $M_z \to -M_z$ which is a
symmetry required by model E.

For the comparison of the critical spin dynamics with the critical dynamics
according to model E it is important to realize, that the configurational
{\em energy} is an {\em additional} constant of motion, because
Eq.(\ref{XYeqmot}), in contrast to the coarse grained model E, does not
contain relaxation. Whether energy conservation is a reasonable assumption
for the dynamics of the XY model or any other classical spin model is a
question of the time scales to be resolved. The most important time scale for
our investigation is set by the propagating modes (spin waves) in the system
and for these the configurational energy is indeed constant. Within the
time scale of the spin waves thermal averages can therefore be replaced by
averages over the initial configurations from which the time integration of
Eq.(\ref{XYeqmot}) is started. For much longer times relaxation processes
(equilibration with the heat bath) come into play which violate energy
conservation and render our spin dynamics approach invalid.
In the vicinity of the critical point energy conservation becomes
particularly important, because the dynamic universality class may change
under the influence of an additional conservation law. If model E is
augmented by energy conservation (model E', see Ref.\cite{HHS76}) it turns
out that the energy asymptotically decouples from the order parameter
$(M_x,M_y)$ and the conserved {\em out-of-plane} magnetization $M_z$ and
model E critical behavior is restored. However, energy conservation may
introduce corrections to the asymptotic finite-size scaling behavior
\cite{HHS76} which decay very slowly and may cause ambiguities in the
scaling analysis of the spin dynamics data. Note that these corrections are
generated by the spin dynamics method.

The equations of motion given by Eq.(\ref{XYeqmot}) are integrated numerically
for each initial spin configuration by a recently developed decomposition
method \cite{spindyn}. This method guarantees exact energy conservation
and conservation of spin length $|{\bf S}_k| = 1$ and conserves $M_z$ within
its numerical truncation errors. For the present study a
second order integrator is used with the time step $\delta t = 0.05/J$.
This time step guarantees sufficient accuracy with respect to the conservation
of $M_z$ and is much faster than well-known predictor corrector methods
\cite{DPLMK98,spindyn}. For some accuracy and stability tests the time
step has been increased to $\delta t = 0.1/J$ which still yields sufficient
accuracy for the dynamic structure factor. Fourth order integrators are
much more accurate as far as $M_z$ conservation is concerned, but their
internal complexity makes them much slower than a second order
method for the same time step \cite{spindyn}. Moreover, statistical errors
are not decrased significantly by fourth order methods and we
therefore only report results obtained by the second order method.
The equations of motion are integrated to a final time of $800/J$ and thermal
averages are taken over 1000 initial configurations. All error bars of
dynamic quantities correspond to one standard deviation. The simulations have
been performed on various DEC alpha AXP, IBM RS6000, and HP RISC8000
workstations both at the RWTH Aachen and the BUGH Wuppertal.

\section{Static properties of the XY model}
\subsection{Thermodynamic properties}
The basic ingredient for the spin dynamics simulation is provided by
the sequence of initial spin configurations, which has to be generated
according to the canonical ensemble in order to provide well defined
thermal averages. Therefore, the static behavior of the XY model
and especially the location of $T_c$
have to be determined first. For this purpose we employ the hybrid
Monte-Carlo scheme described above for lattice sizes $L$ between $L = 20$
and $L = 80$. For each system size and temperature we perform 10 blocks
of $10^3$ hybrid steps for equilibration followed by $10^4$ hybrid steps
for measurements. Each measurement block yields an estimate for all static
quantities of interest and from these we obtain our final estimates and
estimates of their statistical error following standard procedures. The
integrated autocorrelation time of our hybrid algorithm is determined by
the autocorrelation function of the energy or, equivalently, the {\em
modulus} $\sqrt{M_x^2 + M_y^2}$ of the order parameter, which
yield the slowest modes for the
Wolff algorithm. The autocorrelation times are generally rather short,
at $T_c$ (see below) they range from about 5 hybrid MCS
for $L = 20$ to about 10 hybrid MCS for $L = 80$.
For comparison, the autocorrelation function for the order parameter itself
yields an autocorrelation time of less than one hybrid MCS.
The values for the equilibration and measurement periods given above thus
translate to roughly 100 and 1000 autocorrelation times, respectively,
which is sufficient for all practical purposes. In order to obtain the best
statistics for magnetic quantities a measurement is made after every hybrid
MCS.

The critical temperature $T_c$ is determined by monitoring the temperature
and size dependence of the Binder cumulant ratio. Specifically, we measure
the cumulants
\begin{equation}
\label{u1u2}
u_1 = 1 - {\langle M_x^4\rangle \over 3 \langle M_x^2 \rangle^2}
\quad \mbox{and} \quad
u_2 = 1 - {\langle (M_x^2 + M_y^2)^2 \rangle \over 3 \langle M_x^2
+ M_y^2 \rangle^2}.
\end{equation}
It turns out that $u_1$ is more sensitive to changes in temperature and
system size so that we only use $u_1$ for the final fine-tuning of the
temperature $T$. For convenience the temperature is expressed as the
dimensionless reduced coupling $K \equiv J/(k_B T)$, where $K_c \equiv
J/(k_B T_c)$ denotes the critical point. From standard procedures
\cite{CFL93} we obtain $K_c = 0.64440 \pm 0.00005$. For comparison we mention
that $K_c = 0.45420 \pm 0.00002$ for the standard plane rotator XY model on
a simple cubic lattice in $d = 3$ \cite{GH93}. By ignoring corrections to
scaling and averaging over the measurements for $L = 40, 50, 60,$ and $80$
we find the estimates
\begin{equation}
\label{u1cu2c}
u_1^* = 0.3789 \pm 0.0015 \quad \mbox{and} \quad
u_2^* = 0.5859 \pm 0.0008
\end{equation}
for the values $u_1^*$ and $u_2^*$ of the cumulants defined by 
Eq.(\ref{u1u2}) at the critical point. Note that for $K = 0.6444$ the value
of $u_1$ remains within two standard deviations of $u_1^*$ for all $L$. Our
estimate for $u_2^*$ is within two standard
deviations of the corresponding estimate $0.5891 \pm 0.0020$ found in
Ref.\cite{GH93} which already gives some  evidence that the planar Heisenberg
variant of the XY model studied here is indeed a member of the static XY
universality class.

The critical exponents are estimated from the critical finite-size scaling
behavior of the average modulus $\left\langle \sqrt{M_x^2 + M_y^2}
\right\rangle$ of the order parameter, the average square $\langle M_x^2 +
M_y^2 \rangle$ of the order parameter, and the temperature derivative of the
latter. At $T = T_c$, i.e., $K = K_c$ one finds the leading scaling behavior
\begin{eqnarray}
\label{fss}
L^{-3} \left\langle \sqrt{M_x^2 + M_y^2} \right\rangle &\sim&
L^{-\beta / \nu}, \nonumber \\
L^{-3} \left\langle M_x^2 + M_y^2 \right\rangle &\sim&
L^{\gamma / \nu}, \\
{\partial \over \partial T} \ln \left\langle M_x^2 + M_y^2 \right\rangle
&\sim& L^{1 / \nu} \nonumber
\end{eqnarray}
with the system size $L$, where $\beta$, $\gamma$, and $\nu$ are the critical
exponents of the order parameter, the susceptibility, and the correlation
length, respectively. During the data analysis it turns out that corrections
to scaling can be ignored within the statistical error of the quantities in
Eq.(\ref{fss}). From our estimate $K_c = 0.6444$ and $L =$ 20, 24, 30, 36,
40, 50, 60, and 80 we find the following values for the critical
exponents:
\begin{eqnarray}
\label{expon}
\beta / \nu &=& 0.5179 \pm 0.0024, \nonumber \\
\gamma / \nu &=& 1.965 \pm 0.005, \\
1 / \nu &=& 1.494 \pm 0.013 \nonumber.
\end{eqnarray}
These values satisfy the scaling relation $2 \beta/\nu + \gamma/\nu = d$.
Furthermore, we directly obtain from Eq.(\ref{expon}) $\nu = 0.6693 \pm
0.0058$, $\eta = 0.035 \pm 0.005$, $\gamma = 1.315 \pm 0.012$, and $\beta =
0.3467 \pm 0.0034$ in very good agreement with previous Monte-Carlo estimates
\cite{GH93} and renormalization group theory for the $O(N=2)$ Ginzburg-Landau
model \cite{LGZJ85}. The critical exponent $\alpha$ of the specific heat can
be obtained from, e.g. the hyperscaling relation, but the statistical error
of $\nu$ is too large to exclude logarithmic behavior $(\alpha = 0)$.
Apart from this deficiency our simulations confirm XY-like critical
behavior for our version of the XY model (see Eq.(\ref{HXY})) quite
accurately. In the following $k_B T$ is measured in units of $J$ 
chosen such that $k_B T_c = 1$.

\subsection{Static structure factor at criticality}
We continue this section with a short discussion of the static spin-spin
correlation function (structure factor) $G({\bf q})$ at the critical
temperature. The static structure factor is the spatial Fourier transform
of the spin-spin correlation function
\begin{equation}
\label{Gij}
{\cal G}_{\alpha \beta}\left( {\bf R}_i - {\bf R}_j \right)
\equiv \langle S_i^{\alpha} S_j^{\beta} \rangle
- \langle S_i^{\alpha} \rangle \langle S_j^{\beta} \rangle,
\end{equation}
where $\alpha,\beta$ refer to spin the components, ${\bf R}_i$ and ${\bf R}_j$
denote lattice vectors, and the thermal average is indicated by $\langle
\dots \rangle$. Note that ${\cal G}_{\alpha \beta} = {\cal G}_{\alpha \beta}
\left( {\bf R}_i - {\bf R}_j \right) = {\cal G}_{\alpha \beta} \left(
{\bf R}_j - {\bf R}_i \right)$ due to translational invariance and reflection
symmetry of the system. The spatial Fourier transform is therefore given by
a cosine transform which we use in the form
\begin{equation}
\label{Gq}
G_{\alpha \beta}({\bf q}) = \sum_i
{\cal G}_{\alpha \beta}\left( {\bf R}_i \right)\ 
\cos {\bf q} \cdot {\bf R}_i.
\end{equation}
With respect to the spontaneous magnetization of the XY model
$G_{\alpha \beta}({\bf q})$ has a transverse component $G_t({\bf q})$ and
a longitudinal component $G_l({\bf q})$. In our Monte-Carlo simulation we
measure $G_t$ and $G_l$ by {\em rotating} the coordinate system of the spins
around the $z$-axis such that the random but finite magnetization vector
$(M_x,M_y)$ is aligned with the $y$-direction. The $x$- and $y$-components
of the spins in the rotated frame then correspond to the transverse and the
longitudinal spin components, respectively, and their correlation functions
yield $G_t \equiv G_{xx}$ and $G_l \equiv G_{yy}$. It should also be
mentioned that the out-of-plane component $G_{zz}({\bf q})$ of the static
structure factor is independent of ${\bf q}$ and does not show any critical
behavior. According to Eq.(\ref{Gq}) the normalization of
$G_{\alpha \beta}({\bf q})$ is such that $G_{\alpha \beta}({\bf q} = {\bf 0})
= k_B T \chi_{\alpha \beta}$, where $\chi_{\alpha \beta}$ is the static
susceptibility. Note that $G_t({\bf q} = {\bf 0}) = 0$ by definition and
that $G_l({\bf q} = {\bf 0}) = k_B T \chi'$, where $\chi'$ is the magnetic
susceptibility with respect to the {\em modulus} of the magnetization. We
furthermore limit the discussion to the $(100)$ direction, because other
lattice directions do not provide new information on a simple cubic lattice.
The components of ${\bf q}$ are always measured in units of the inverse
lattice contant.

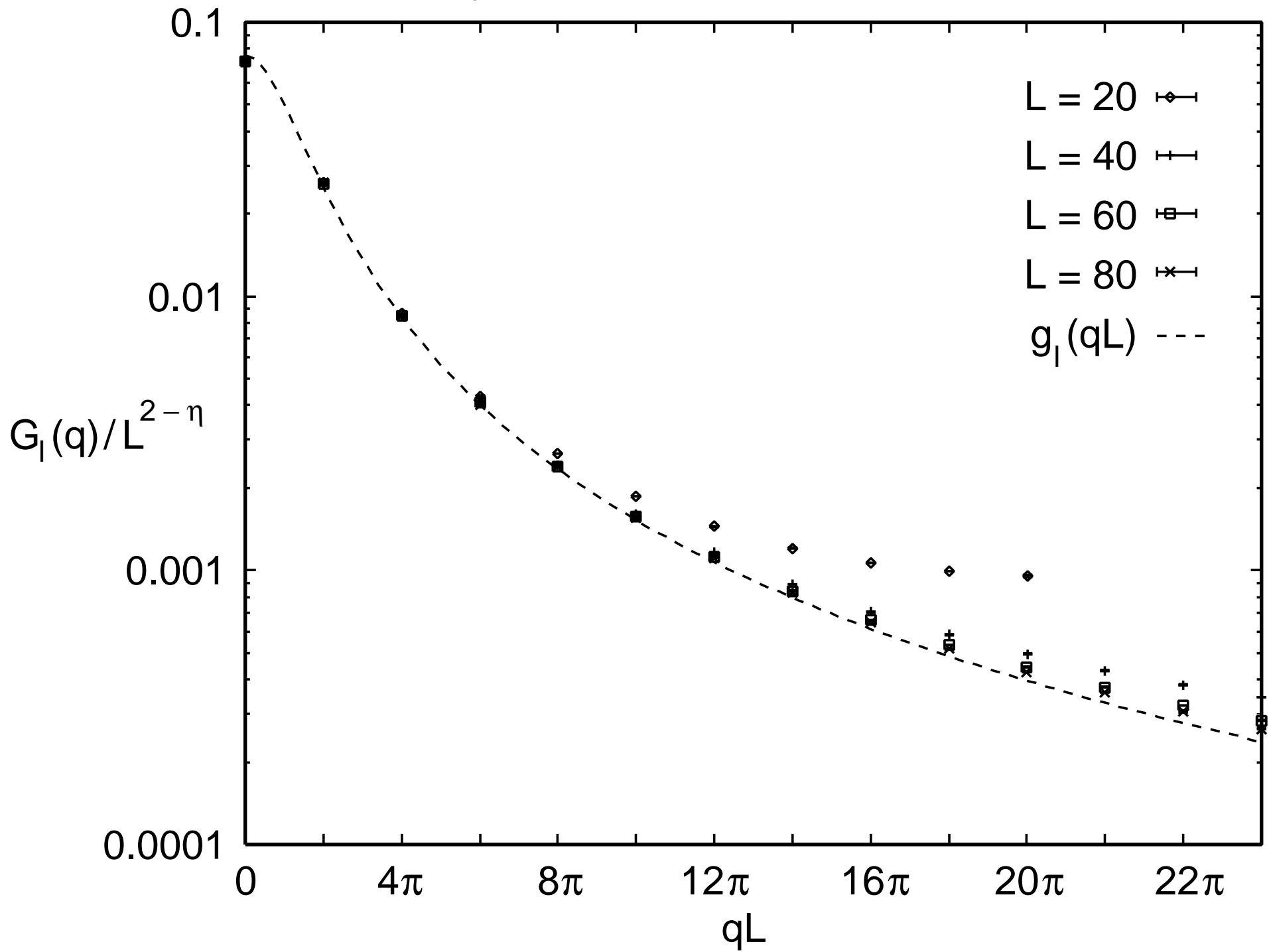
\begin{figure}[t]
\caption{Scaling plot for $G_l({\bf q})/L^{2-\eta}$ at $T = T_c$ for
$L = 20$ ($\Diamond$), 40 (+), 60 ($\Box$), and 80 ($\times$) as a function
of $qL$ with $\eta = 0.035$. Statistical errors of the data are much smaller
than the symbol sizes. The dashed line displays the scaling function $g_l(qL)$
as given by Eq.(\protect\ref{glx}) for $a = 3.70$ and $b = 4.35$. Note that
$qL = 20\pi$ corresponds to the Brillouin zone boundary for $L = 20$.
\label{Glscal}}
\end{figure}

At $T_c$ the longitudinal component $G_l({\bf q})$ of the static structure
factor can be described by the model function
\begin{equation}
\label{Glq}
G_l\left( {\bf q} = (q,0,0) \right) = L^{2-\eta} g_l(qL) h(q),
\end{equation}
where $2 - \eta = \gamma/\nu$ is taken from Eq.(\ref{expon}), $h(q) =
[(q/2)/\sin(q/2)]^2$ captures lattice effects \cite{RF72}, and the
finite-size scaling function $g_l(x)$ is chosen as a simple generalization
of a Lorentzian:
\begin{equation}
\label{glx}
g_l(x) = a^{-(2-\eta)} \left[1 + (x/b)^2 \right]^{-(2-\eta)/2} .
\end{equation}
The parameters $a$ and $b$ are determined from a fit to the data, where
$a$ and $b$ are independent of $L$. We have chosen $a = 3.70$ and $b = 4.35$
as obtained from a fit to the data for $L = 60$. For $L = 40$ and $L = 80$
$a$ and $b$ are found within less than 1\% of these values, for $L = 20$ they
are about 3\% smaller. Despite its simplicity the
model function given by Eq.(\ref{Glq}) captures the shape of the Monte-Carlo
estimate for the longitudinal component of the structure factor remarkably
well. The finite-size scaling analysis of our data
for $G_l({\bf q})$ for small $q$, is shown in Fig.\ref{Glscal}.
Deviations from finite-size scaling set in at $qL = 4\pi$ for $L = 20$,
the data for $L = 40$, 60, and 80 collapse within the error bars up to
$qL = 6\pi$. Within the symbol sizes data collapse is obtained up to
$qL = 10\pi$, where lattice effects set in. For comparison the model
scaling function $g_l(qL)$ ($a = 3.70$ and $b = 4.35$, see Eq.(\ref{glx}))
is shown by the dashed line in Fig.\ref{Glscal}.
The true scaling form of $G_l({\bf q})$ for small $q$ is captured rather
well by $g_l(x)$. However, the choice of the model function is, of course,
not unique.

\section{The dynamic structure factor}
The dynamic structure factor $S({\bf q},\omega)$ is the space-time
Fourier transform of the position and time displaced spin-spin correlation
function ${\cal S}_{\alpha \beta} \left( {\bf R}_i - {\bf R}_j,|t - t'|
\right)$ which we define by (see Eq.(\ref{Gij}))
\begin{equation}
\label{Sijtt}
{\cal S}_{\alpha \beta}\left( {\bf R}_i - {\bf R}_j, |t - t'| \right)
\equiv \langle S_i^{\alpha}(t) S_j^{\beta}(t') \rangle 
- \langle S_i^{\alpha}(t) \rangle \langle S_j^{\beta}(t') \rangle .
\end{equation}
The indices $\alpha, \beta$ refer to spin components, ${\bf R}_i$ and
${\bf R}_j$ are lattice vectors, and $t$ and $t'$ are moments in time
to which the initial spin configuration has evolved according to the
equations of motion (see Eq.(\ref{XYeqmot})). The average $\langle \dots
\rangle$ is taken over the set of initial configurations as described in
Sec.II. Note that ${\cal S}_{\alpha \beta}$ is also symmetric with respect
to ${\bf R}_i$ and ${\bf R}_j$ (see Eq.(\ref{Gij})). As the second argument
only the time displacement enters, because the equations of motion are
invariant under the transformation ${\bf S}_i \to -{\bf S}_i$, $t \to -t$
(see Eqs.(\ref{HXY}) and (\ref{XYeqmot})). The space-time Fourier transform
is therefore also given by the cosine transform (see Eq.(\ref{Gq}))
\begin{equation}
\label{Sqw}
S_{\alpha \beta}({\bf q},\omega) = 2/\pi \int_0^{\infty} \sum_i
{\cal S}_{\alpha \beta}\left( {\bf R}_i, |t| \right)\ 
\cos {\bf q} \cdot {\bf R}_i\ \cos \omega t\ dt.
\end{equation}
Note that the normalization in Eq.(\ref{Sqw}) is such that $\int_0^{\infty}
S_{\alpha \beta}({\bf q},\omega) d\omega = G_{\alpha \beta}({\bf q})$, where
$G_{\alpha \beta}({\bf q})$ is the static structure factor defined by
Eq.(\ref{Gq}).

The out-of-plane component $S_{zz}({\bf q},\omega)$ of the dynamic structure
factor is associated with the conserved out-of-plane magnetization $M_z$,
i.e., $\sum_i {\cal S}_{zz}\left({\bf R}_i, |t| \right)$ (see also
Eq.(\ref{Sqw})) is a constant in time. The in-plane magnetization
(order parameter) $(M_x,M_y)$ is not conserved under the spin dynamics.
Although the time scale set by the motion of $M_x$ and $M_y$
is considerably larger than typical time scales set by spin waves, all
initial differences between longitudinal and transverse components of the
spin correlations completely disappear during the integration of the
equations of motion. We therefore only discuss the average of the $xx$ and
the $yy$ component of $S({\bf q},\omega)$ and refer to it as the "in-plane
component" $S_{xx}({\bf q},\omega)$.

We now turn to the discussion of $S({\bf q},\omega)$ above, below, and at
the critical point. The correlation functions are measured up to a time
displacement of $400 / J$ away from criticality, where only smaller systems
are considered. At the critical temperature we measure correlations up to
$600 / J$ for systems with up to $L = 60$ lattice sites in each direction.

\subsection{The structure factor above $T_c$}
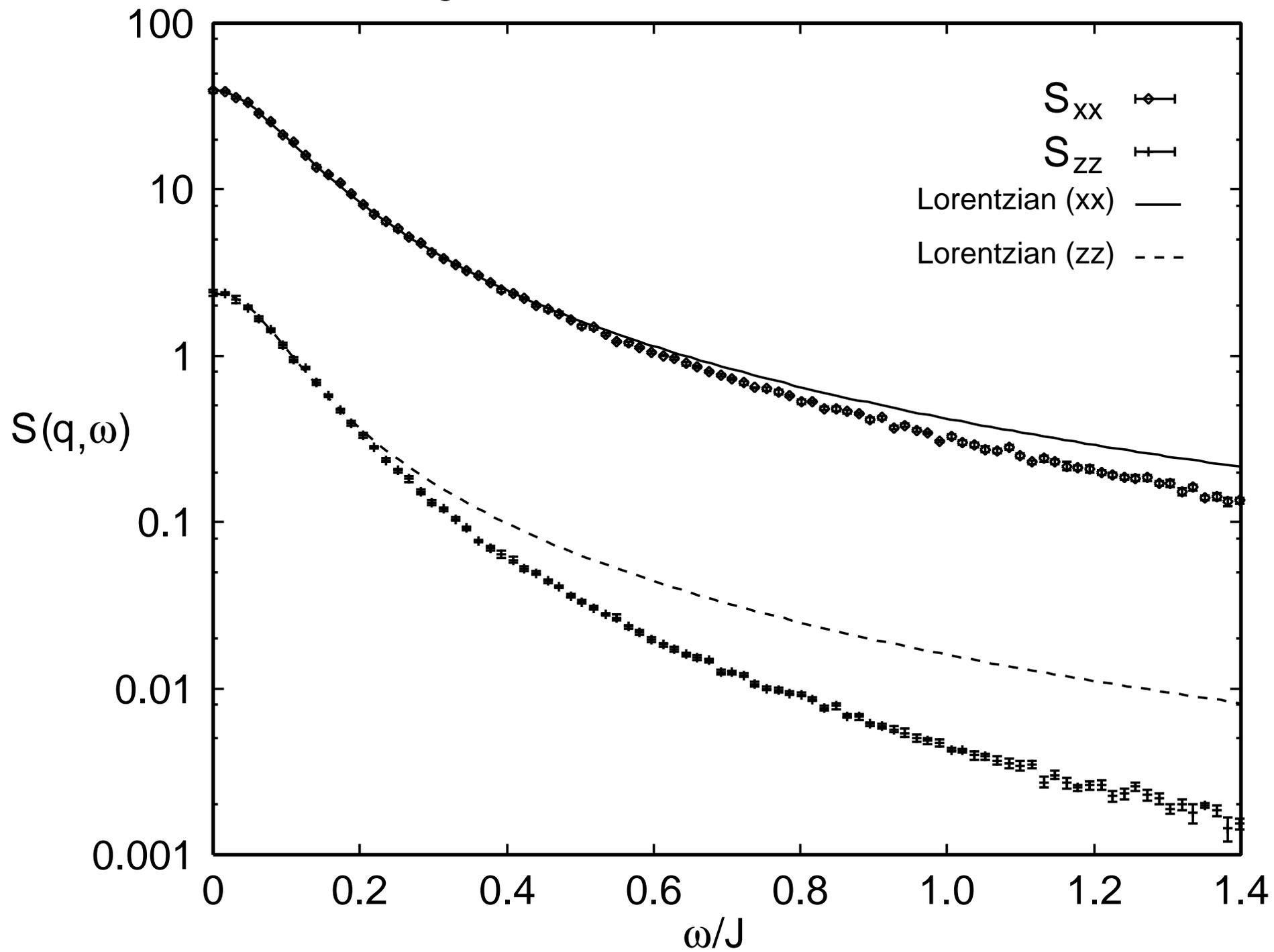
\begin{figure}[t]
\caption{In-plane component $S_{xx}$ ($\Diamond$) and out-of-plane component
$S_{zz}$ (+) of the dynamic structure factor in the $(100)$ direction for
$T = 1.1 T_c$, $L = 30$, and $q = \pi/15$ as functions of the dimensionless
frequency $\omega / J$. For small enough $\omega$ the line shapes are well
approximated by Lorentzians given by Eq.(\protect\ref{Lxx}) (solid
line) and Eq.(\protect\ref{Lzz}) (dashed line), respectively. The ranges
over which the Lorentzian approximation is valid differ significantly for
$S_{xx}$ and $S_{zz}$.
\label{Sxzfit}}
\end{figure}

In order to avoid effects of criticality we choose the temperature
$T = 1.1 T_c$ in the following. Due to the absence of critical finite-size
scaling at this temperature we limit the spin dynamics simulations
to smaller systems with $L = 20$, 24, and 30. For
better momentum resolution we only present results obtained for $L = 30$,
smaller $L$ yield identical results with a lower resolution in ${\bf q}$.
We also limit the presentation to $S({\bf q},\omega)$ in the $(100)$
direction, other lattice directions provide essentially the same
information. In Fig.\ref{Sxzfit} we show $S_{xx}({\bf q}=(q,0,0),\omega)$
and $S_{zz}({\bf q}=(q,0,0),\omega)$ for $q = \pi/15$ as functions of
the dimensionless frequency $\omega / J$. Both components apparently display
a central peak without any additional features. For the values of
$T$ and $q$ used in Fig.\ref{Sxzfit} one expects a central peak
of Lorentzian shape \cite{TFS91}. As displayed by the solid line in
Fig.\ref{Sxzfit} a simple Lorentzian of the form
\begin{equation}
\label{Lxx}
L_{xx}(q,\omega) = {A^0_{xx}(q) \over 1 + [\omega / \Gamma^0_{xx}(q)]^2}
\end{equation}
characterized by an amplitude $A^0_{xx}(q)$ and a width $\Gamma^0_{xx}(q)$
captures the shape of $S_{xx}$ very well up to $\omega / J \simeq 0.4$,
where the intensity has already dropped by an order of magnitude. For larger
$\omega$ $S_{xx}$ decays faster than a Lorentzian. If one tries to fit the
line shape of $S_{zz}$ also with a Lorentzian for small $\omega$, it turns
out that a better fit is obtained from a superposition of symmetrically
placed Lorentzians. We use
\begin{equation}
\label{Lzz}
L_{zz}(q,\omega) =
{A_{zz}(q) \over 1 + \left[(\omega - \omega(q)) / \Gamma_{zz}(q) \right]^2} +
{A_{zz}(q) \over 1 + \left[(\omega + \omega(q)) / \Gamma_{zz}(q) \right]^2} ,
\end{equation}
where $\omega(q)$ denotes the spin wave frequency which is used as an
additional fit parameter. The dashed line in
Fig.\ref{Sxzfit} displays the Lorentzian fit for $S_{zz}$ according to
Eq.(\ref{Lzz}), where $\omega(\pi/15) \simeq 0.030$ is indeed finite within
its statistical error of roughly $10^{-3}$. However, Eq.(\ref{Lzz}) only
captures the shape of $S_{zz}$ up to $\omega / J \simeq 0.15$. If
Eq.(\ref{Lxx}) ($\omega(\pi/15) = 0$) is used instead, this
frequency range becomes even smaller. This discrepancy in frequency range
between the in-plane and the out-of-plane components of $S({\bf q},\omega)$
may be due to the difference in time scales between in-plane and
out-of-plane modes as predicted by mode coupling theory \cite{TFS91}.

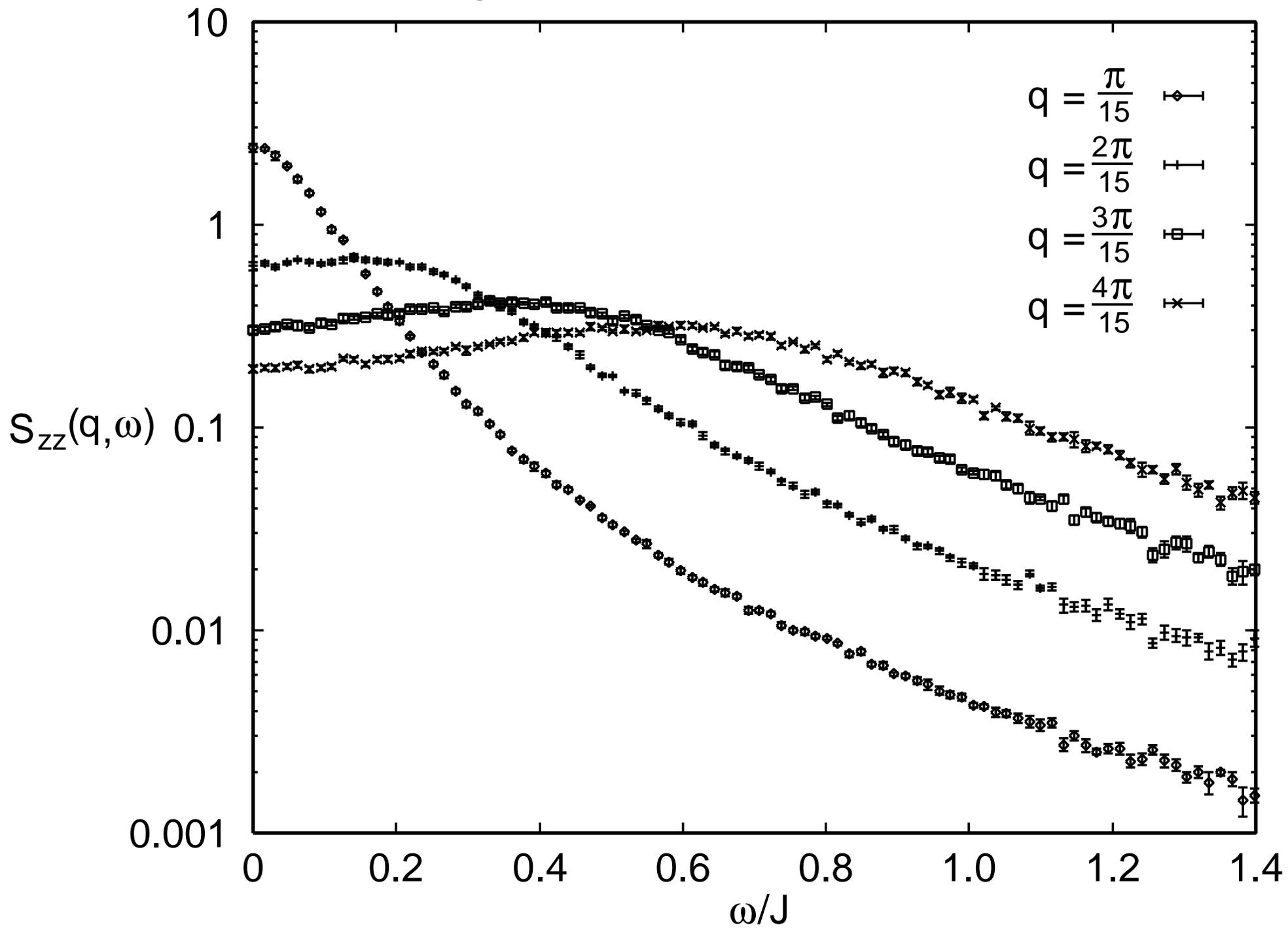
\begin{figure}[t]
\caption{Out-of-plane component $S_{zz}$ of the dynamic structure factor
in the $(100)$ direction for $T = 1.1 T_c$, $L = 30$, and $q = \pi/15$
($\Diamond$), $q = 2\pi/15$ (+), $q = 3\pi/15$ ($\Box$), $q = 4\pi/15$
($\times$) as a function of the dimensionless frequency $\omega / J$.
\label{SzzaTc}}
\end{figure}

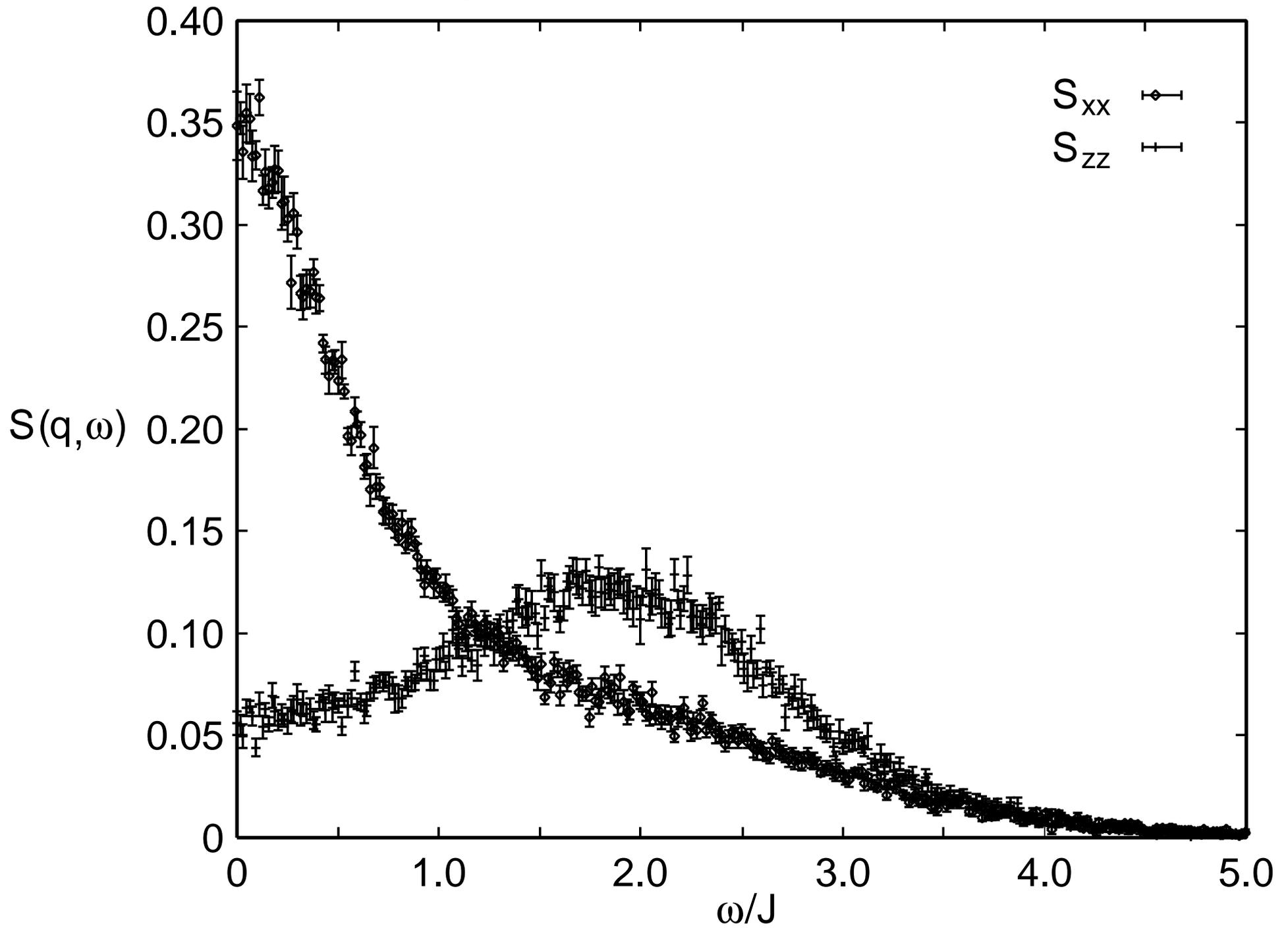
\begin{figure}[t]
\caption{In-plane component $S_{xx}$ ($\Diamond$) and out-of-plane component
$S_{zz}$ (+) of the dynamic structure factor in the $(100)$ direction for
$T = 1.1 T_c$, $L = 30$, and $q = \pi$ as functions of the dimensionless
frequency $\omega / J$. A shoulder appears at the position of
the spin wave peak in $S_{zz}$.
\label{Sxzpi}}
\end{figure}

Spin wave signatures are still visible in $S_{zz}$ at $T = 1.1 T_c$ as shown
in Fig.\ref{SzzaTc}, where $S_{zz}$ is plotted for the first four momenta
$q = n \pi/15$, $n = 1,2,3,4$. For $n \geq 2$ a very broad maximum becomes
visible which moves to the right as $q$ increases. The appearance of spin
wave signatures in $S_{zz}$ above $T_c$ is expected from renormalization
group theory for model E dynamics \cite{Dohm78}. For the $q$ values used
in Fig.\ref{SzzaTc} the shape of $S_{zz}$ is captured quite well by the
Lorentzian defined by Eq.(\ref{Lzz}). Due to the small enhancement of the
line intensity over the signal level at $\omega = 0$ the peak position
$\omega(q)$ and linewidth $\Gamma_{zz}(q)$ are not very well defined. We
therefore refrain from a more detailed analysis at this point. The frequency
dependence of the in-plane component $S_{xx}$ is dominated by a central peak
as in Fig.\ref{Sxzfit} for all $q$. Only for large $q$ near the Brillouin
zone boundary a shoulder appears in $S_{xx}$ near the position of the
spin-wave peak in $S_{zz}$. We illustrate this in Fig.\ref{Sxzpi}, where
$S_{xx}$ and $S_{zz}$ are shown for $q = \pi$. Note that $S_{zz}$, which
is much smaller than $S_{xx}$ for small $q$ (see Fig.\ref{Sxzfit}), becomes
comparable to $S_{xx}$ in magnitude for large $q$. In a qualitative sense
we have recovered the same behavior as observed by spin dynamics simulations
of the XY model in $d = 2$ above the Kosterlitz - Thouless transition
\cite{EvLan96}.

\subsection{The structure factor below $T_c$}
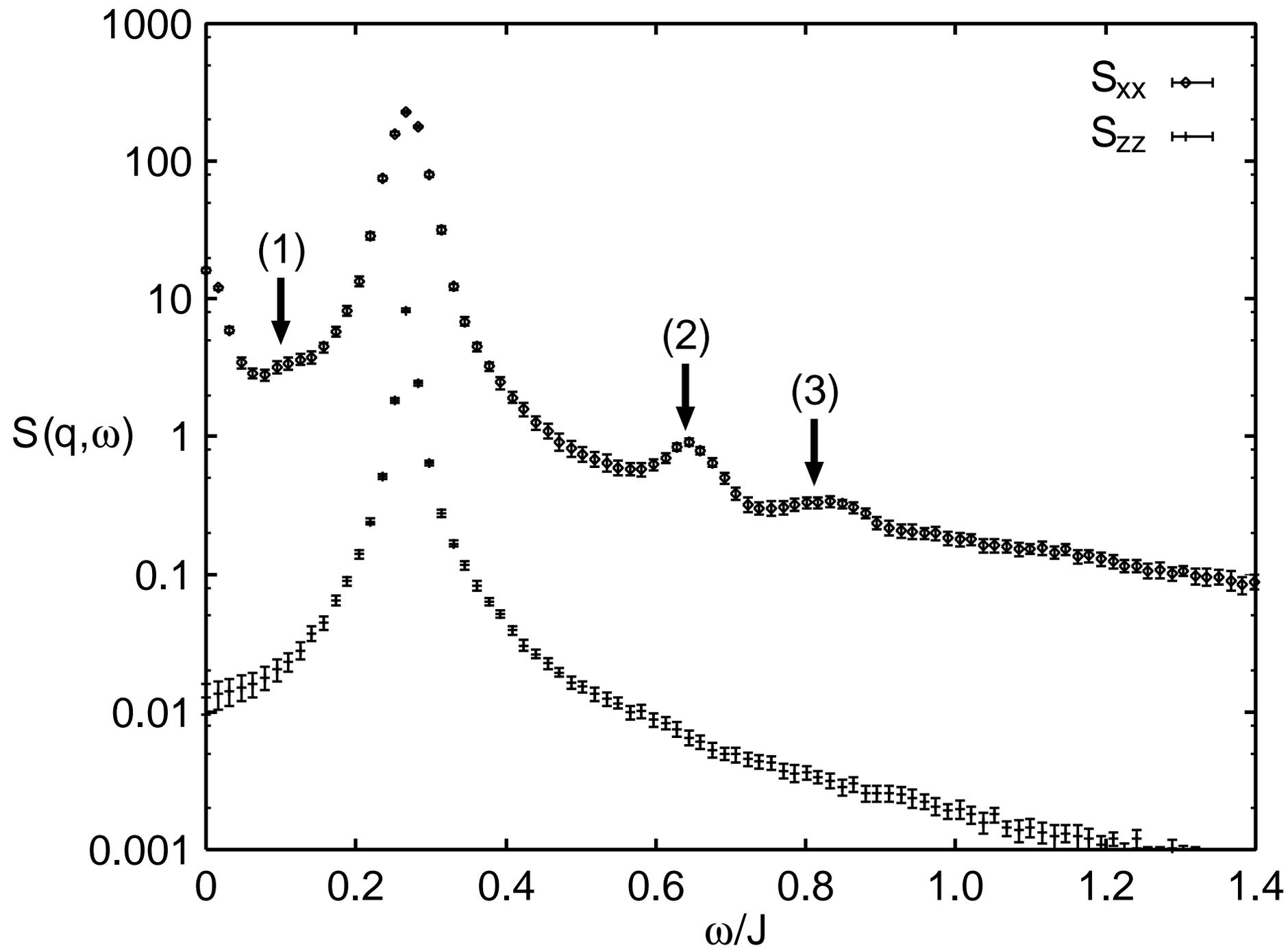
\begin{figure}[t]
\caption{In-plane component $S_{xx}$ ($\Diamond$) and out-of-plane component
$S_{zz}$ (+) of the dynamic structure factor in the $(100)$ direction for
$T = 0.9 T_c$, $L = 30$, and $q = \pi/15$ as functions of the dimensionless
frequency $\omega / J$. Apart from the dominant spin wave peak additional
resonances appear in $S_{xx}$ (arrows) at low intensities (see main text).
\label{SxzSW}}
\end{figure}

As before we want to avoid critical finite-size effects also below $T_c$
and we therefore choose $T = 0.9 T_c$ in the following. Again relatively
small systems are sufficient for this investigation. Our main results
have been obtained for $L = 30$, smaller systems yield the same results
at a lower $q$ resolution. The dynamics of the XY model below $T_c$ is
dominated by spin waves which are visible in $S({\bf q},\omega)$ as
pronounced peaks at the spin wave frequency $\omega(q)$ as shown in
Fig.\ref{SxzSW} for $q = \pi/15$ in the $(100)$ direction. The peak
intensity of $S_{xx}$ is more than one order of magnitude larger than
the corresponding peak intensity of $S_{zz}$. $S_{xx}$ also
displays a pronounced central peak, where the intensity exceeds the
intensity of $S_{zz}$ by over three orders of magnitude. In a qualitative
sense this is again the same behavior as observed in $d = 2$ \cite{EvLan96},
where the central peak in $S_{xx}$ was not expected by analytical theory
\cite{Mertens}. We also find some 'fine structure' at low intensities in
$S_{xx}$ for which no theoretical predictions exist. In Fig.\ref{SxzSW} we
have marked these additional resonances by arrows. As demontrated already
in $d = 2$ \cite{EvLan96} these signals can be interpreted as two-spin wave
peaks in the following way. In order to produce a contribution to $S_{xx}$
at $q = \pi/15$ in the $(100)$ direction one can combine {\em two} spin waves
at $q = \pi/15$, one in the $(010)$ direction, which is eqivalent to the
$(100)$ direction on a simple cubic lattice, and one in the $(110)$
direction (not shown), where the latter has a higher energy (frequency).
The {\em difference} between the corresponding frequencies is marked as
$(1)$ in Fig.\ref{SxzSW}, their {\em sum} is marked as $(2)$. A second way
to get a contribution to $S_{xx}$ at $q = \pi/15$ in the $(100)$ direction
is to combine two spin waves in the $(100)$ direction, one at $q = \pi/15$
the other at $q = 2\pi/15$, where again the latter has a higher energy.
The {\em sum} of the corresponding frequencies is marked as $(3)$ in
Fig.\ref{SxzSW} their {\em difference} almost coincides with the position
of the spin wave peak and can therefore not be resolved.

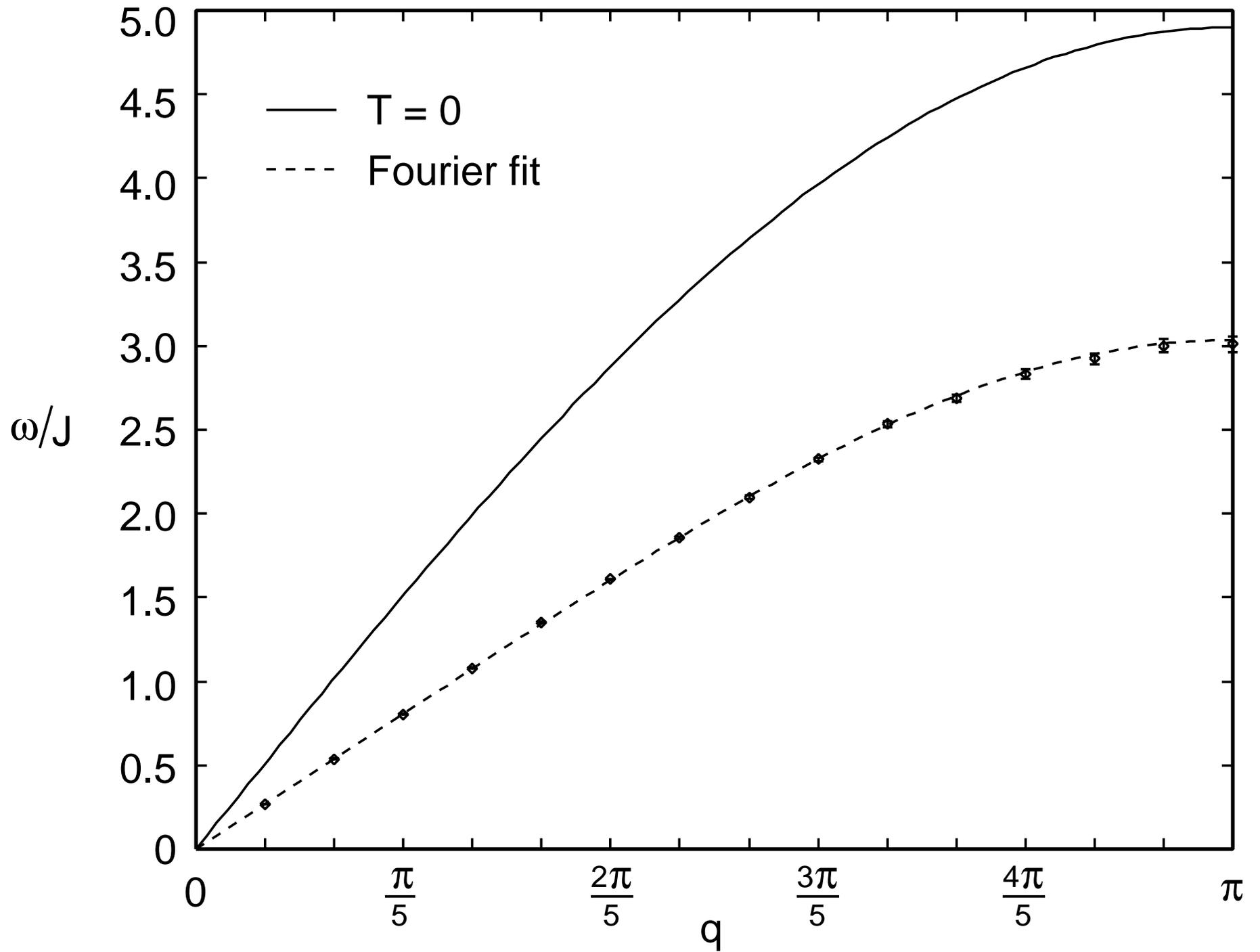
\begin{figure}[t]
\caption{Dispersion relation $\omega(q)$ ($\Diamond$) in the $(100)$ direction
for $T = 0.9 T_c$ and $L = 30$. The solid line is the $T = 0$ dispersion
relation obtained from linear spin wave theory (see Eq.(\protect\ref{omqT0})).
The dashed line is a fit to the first two terms of the Fourier series given
by Eq.(\protect\ref{omqFour}).
\label{omq}}
\end{figure}

In order to monitor the $q$ dependence of the spin wave frequency
$\omega(q)$ and the line widths $\Gamma_{xx}(q)$ and $\Gamma_{zz}(q)$
of the spin wave peaks, we again employ fits to simple Lorentzians. For
the in-plane component we generalize Eq.(\ref{Lxx}) to include the spin
wave peaks:
\begin{equation}
\label{LxxSW}
L_{xx}(q,\omega) = {A^0_{xx}(q) \over 1 + [\omega / \Gamma^0_{xx}(q)]^2}
+ {A_{xx}(q) \over 1 + \left[(\omega - \omega(q)) / \Gamma_{xx}(q) \right]^2}
+ {A_{xx}(q) \over 1 + \left[(\omega + \omega(q)) / \Gamma_{xx}(q) \right]^2}.
\end{equation}
For the out-of-plane component we use Eq.(\ref{Lzz}) which captures the
shape of $S_{zz}$ for sufficiently small frequencies
in a satisfactory way. The dispersion relation $\omega(q)$
can be obtained from Eq.(\ref{LxxSW}) or Eq.(\ref{Lzz}), where the latter
yields slightly smaller error bars, because $S_{zz}$ displays a sharper
spin wave maximum. The estimates for $\omega(q)$, $\Gamma_{xx}(q)$, and
$\Gamma_{zz}(q)$ depend on the frequency range over which Eqs.(\ref{LxxSW})
and (\ref{Lzz}) are fitted to $S_{xx}$ and $S_{zz}$, respectively. We have
chosen a frequency window around the spin wave peak, where for $S_{xx}$
the central peak has been subtracted first. The error in the dispersion
relation and the line widths is estimated by varying the size of the
frequency window from about 1.2 times the half width to about twice the
half width of the peak. The dispersion relation $\omega(q)$ obtained
from this procedure for the $(100)$ direction is shown in Fig.\ref{omq},
the correponding zero temperature dispersion relation shown by the solid
line. According to linear spin wave theory one obtains
\begin{equation}
\label{omqT0}
\omega(q) / J = 2 \sqrt{2d} \sin(q/2) ,
\end{equation}
for $T = 0$ where $d = 3$ in our case. As expected, the spin wave frequencies
are 'renormalized' to fall below the $T = 0$ dispersion curve. The functional
form of $\omega(q)$ can be captured by a Fourier series. A convenient choice
is
\begin{equation}
\label{omqFour}
\omega(q) / J = a_1 \sin(q/2) + a_2 \sin(3q/2) + a_3 \sin(5q/2) + \dots ,
\end{equation}
where the Fourier coefficients $a_1, a_2, a_3, \dots$ are rapidly decreasing.
If Eq.(\ref{omqFour}) is truncated after the second term one obtains $a_1 =
2.919$ and $a_2 = -0.116$ from a least square fit which
connects all data points within their error bars as shown by the dashed
line in Fig.\ref{omq}.

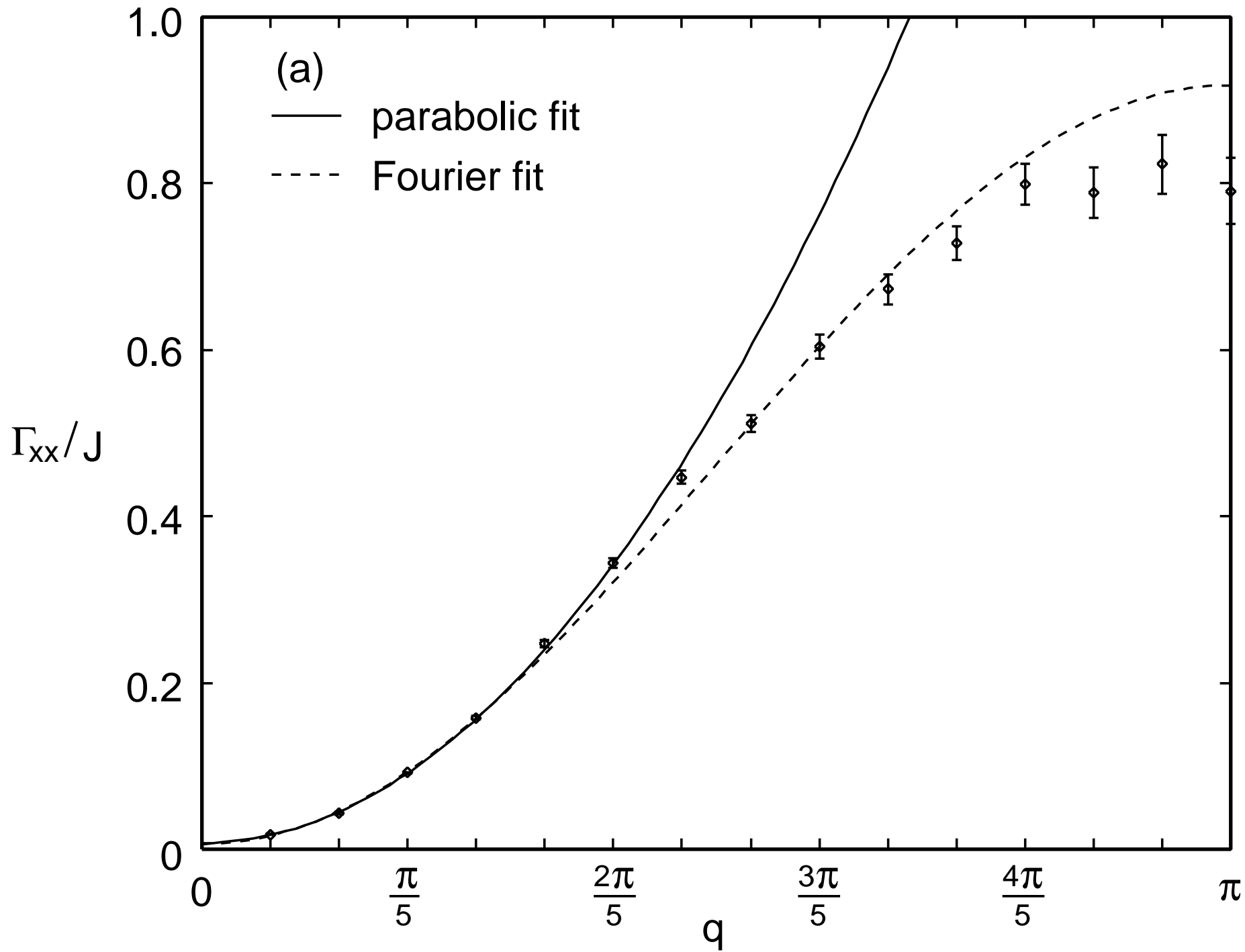
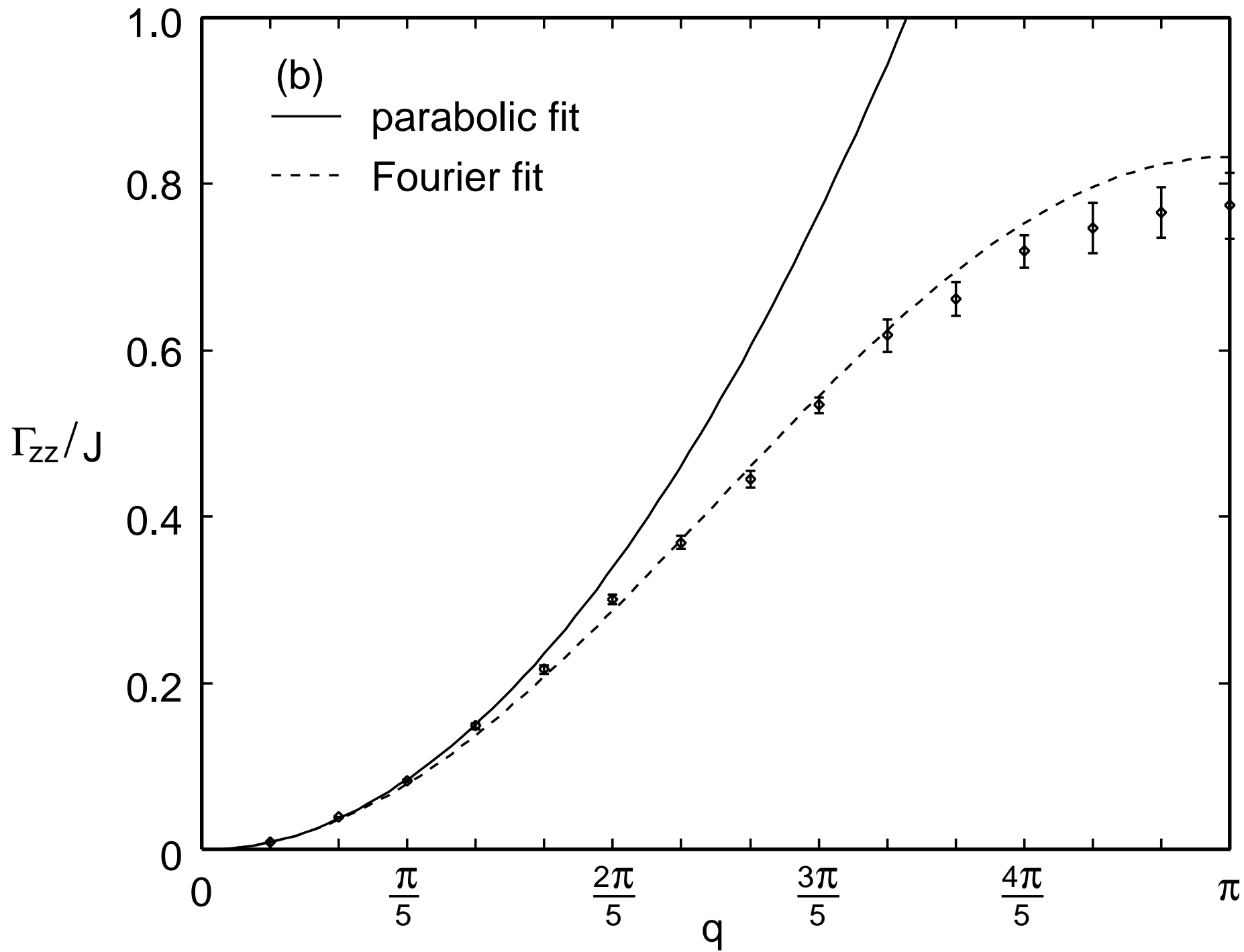
\begin{figure}[t]
\caption{Line widths (a) $\Gamma_{xx}(q)$ and (b) $\Gamma_{zz}(q)$ of the
spin wave peak in $S_{xx}$ for the $(100)$ direction, $T = 0.9 T_c$, and
$L = 30$. The solid lines are parabolic fits to the first four data
points (see Fig.\protect\ref{Gamq2}). The dashed lines are fits with the
first two terms of the Fourier series given by Eq.(\protect\ref{GamFour}).
\label{Gxxq}}
\end{figure}

The line width $\Gamma_{xx}(q)$ of the spin wave peak in the in-plane
component of the structure factor is shown in Fig.\ref{Gxxq}(a). As before
the data can be analyzed by a Fourier series with rapidly decreasing
coefficients. A convenient choice here is
\begin{equation}
\label{GamFour}
\Gamma_{\alpha \alpha}(q) / J = b_0 + b_1 \cos q + b_2 \cos 2q + \dots ,
\end{equation}
where $\alpha$ refers to $x$ (in-plane) and $z$ (out-of-plane). A fit
to $\Gamma_{xx}$ with only two coefficients
yields $b_0 = 0.4622$ and $b_1 = -0.4559$ which is shown by the
dashed line in Fig.\ref{Gxxq}(a). The statistical error of $b_0$ and $b_1$
is about $5 \times 10^{-4}$. The data for small $q$ are represented
quite well, whereas near the Brillouin zone boundary significant deviations
occur. These can be reduced very quickly by including higher Fourier modes
in the fit (not shown), where the higher Fourier coefficients decrease
rapidly in magnitude. Here, we are primarily interested in the $q$ dependence
of the line width near the center of the Brillouin zone (see below). A
corresponding analysis has been
performed for the line width $\Gamma_{zz}(q)$ of the spin wave peak in the
out-of-plane component of the structure factor. The result is shown in
Fig.\ref{Gxxq}(b), where the Fourier coefficients in the fit (dashed line)
are given by $b_0 = 0.4160(5)$ and $b_1 = -0.4165(5)$. Again the small $q$
behavior is captured very well by the fit, but near the Brillouin zone
boundary deviations occur which can also be reduced very quickly by
including higher Fourier modes.

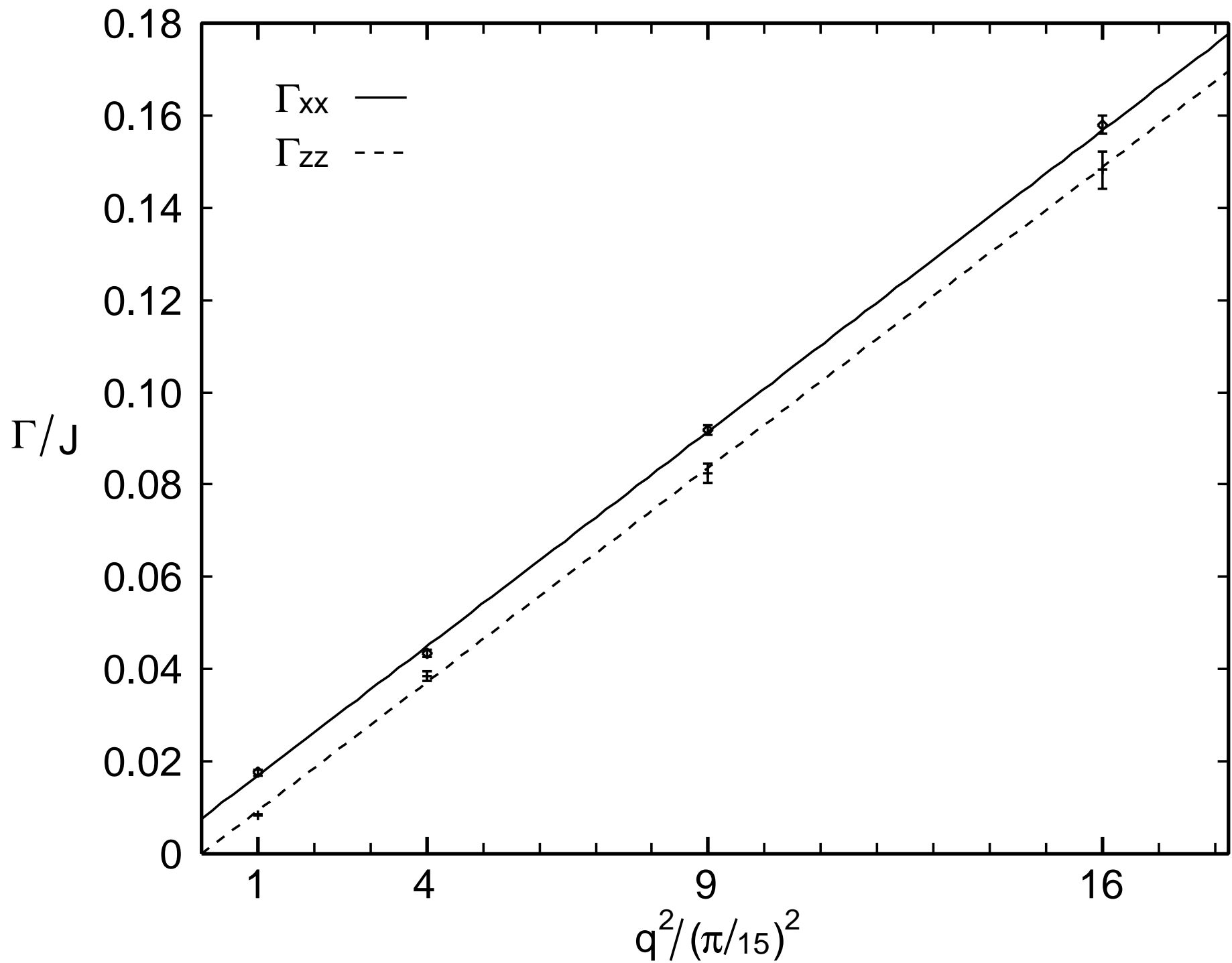
\begin{figure}[t]
\caption{Extrapolation of the line widths $\Gamma_{xx}(q)$ (solid line)
and $\Gamma_{zz}(q)$ (dashed line) to $q = 0$. The line widths vary as
$q^2$ for small $q$ as predicted by mode coupling theory (see main text).
\label{Gamq2}}
\end{figure}

The limit $q \to 0$ is of particular interest for the line widths,
because it reflects the influence of conservation laws on the dynamics.
The line width $\Gamma_{xx}(q=0)$ can be interpreted as the relaxation rate
of the (non-conserved) order parameter and therefore $\Gamma_{xx}(q=0)$
should be positive. The line width $\Gamma_{zz}(q=0)$ is the relaxation
rate of the out-of-plane magnetization $M_z$ which is {\em conserved},
i.e., $\Gamma_{zz}(q=0)$ should {\em vanish}.
From the Fourier fits shown in Fig.\ref{Gxxq} one
finds $\Gamma_{xx}(0) \simeq 0.0063$ and $\Gamma_{zz}(0) \simeq -0.0005$
with statistical errors of about $7 \times 10^{-4}$ in both cases.
The extrapolation of $\Gamma_{xx}(q)$ and $\Gamma_{zz}(q)$ to $q = 0$ is
shown in Fig.\ref{Gamq2}. The $q$ dependence of the line widths for small
$q$ is quadratic as anticipated by the Fourier fits shown in Fig.\ref{Gxxq}.
From a fit to a straight line we find $\Gamma_{xx}(0) = 0.0076 \pm 0.0006
> 0$ (solid line), whereas $\Gamma_{zz}(0) = 0$ (dashed line) within its
statistical error in agreement with the conservation laws. Finally, we
note that the $q^2$ dependence of the linewidth $\Gamma_{zz}$ is in
agreement with the prediction of mode coupling theory \cite{TFS91}.

\subsection{The structure factor at $T_c$}
The exploration of critical dynamics and dynamic scaling naturally requires
most of the numerical effort. In order to reach the scaling regime large
systems are required and we have therefore performed simulations for $L =$
20, 24, 30, 40, and 60. Our prime objective here is the test of dynamic
finite-size scaling which we assume to be valid in the form
\cite{CL94,EvLan96}
\begin{equation}
\label{Sqwscal}
S_{\alpha \alpha}({\bf q},\omega) / G_{\alpha \alpha}({\bf q}) =
L^{z_{\alpha}} \Sigma_{\alpha \alpha}(qL, \omega L^{z_{\alpha}})
\end{equation}
at the critical point, where $G_{\alpha \alpha}$ is the static structure
factor discussed in Sec.III and $z_{\alpha}$ denotes the dynamic critical
exponent. Note, that the ${\bf q}$ dependence is reduced to a dependence
only on $q = |{\bf q}|$, i.e., isotropic scaling in space has been assumed.
The index $\alpha$ again refers to the spin component $x$ (in-plane) or $z$
(out-of-plane), where $z_x \equiv z_{\phi}$ and $z_z \equiv z_m$, respectively.
In order to estimate $z_{\alpha}$ we scale our data according to
Eq.(\ref{Sqwscal}) and test the result for data collapse in the limit of
small frequencies. The static structure factor needed for normalization in
Eq.(\ref{Sqwscal}) is given by the equal-time correlations which we also
extract from the spin dynamics data for consistency.

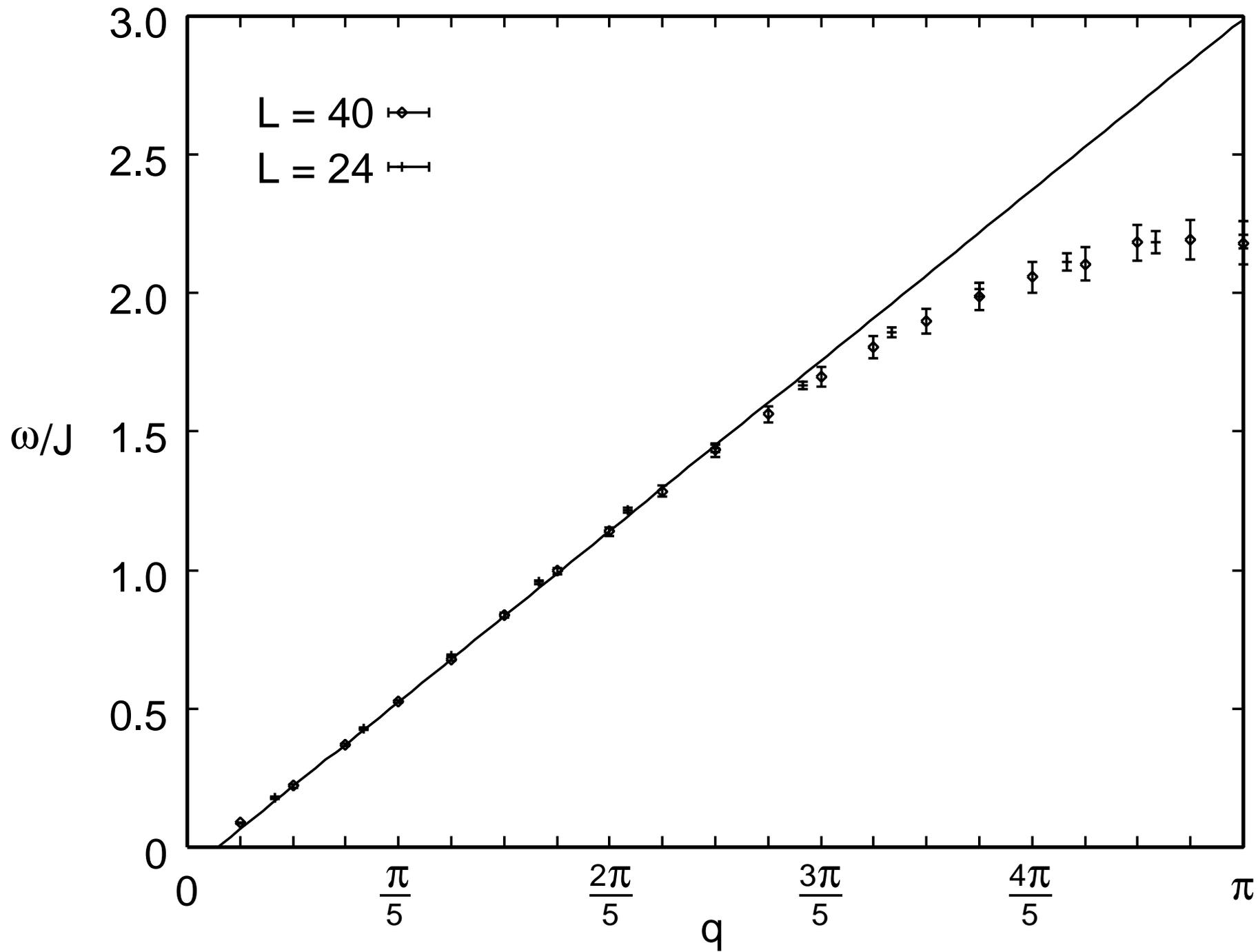
\begin{figure}[t]
\caption{Dispersion relation $\omega_{zz}(q)$ in the $(100)$ direction for
$T = T_c$ and $L = 24$ (+) and 40 ($\Diamond$). The solid line displays a
linear extrapolation of the data for $q < \pi/2$ to $q = 0$. Deviations from
linearity become visible only for $q \leq \pi/20$ where critical effects
set in.
\label{omTc}}
\end{figure}

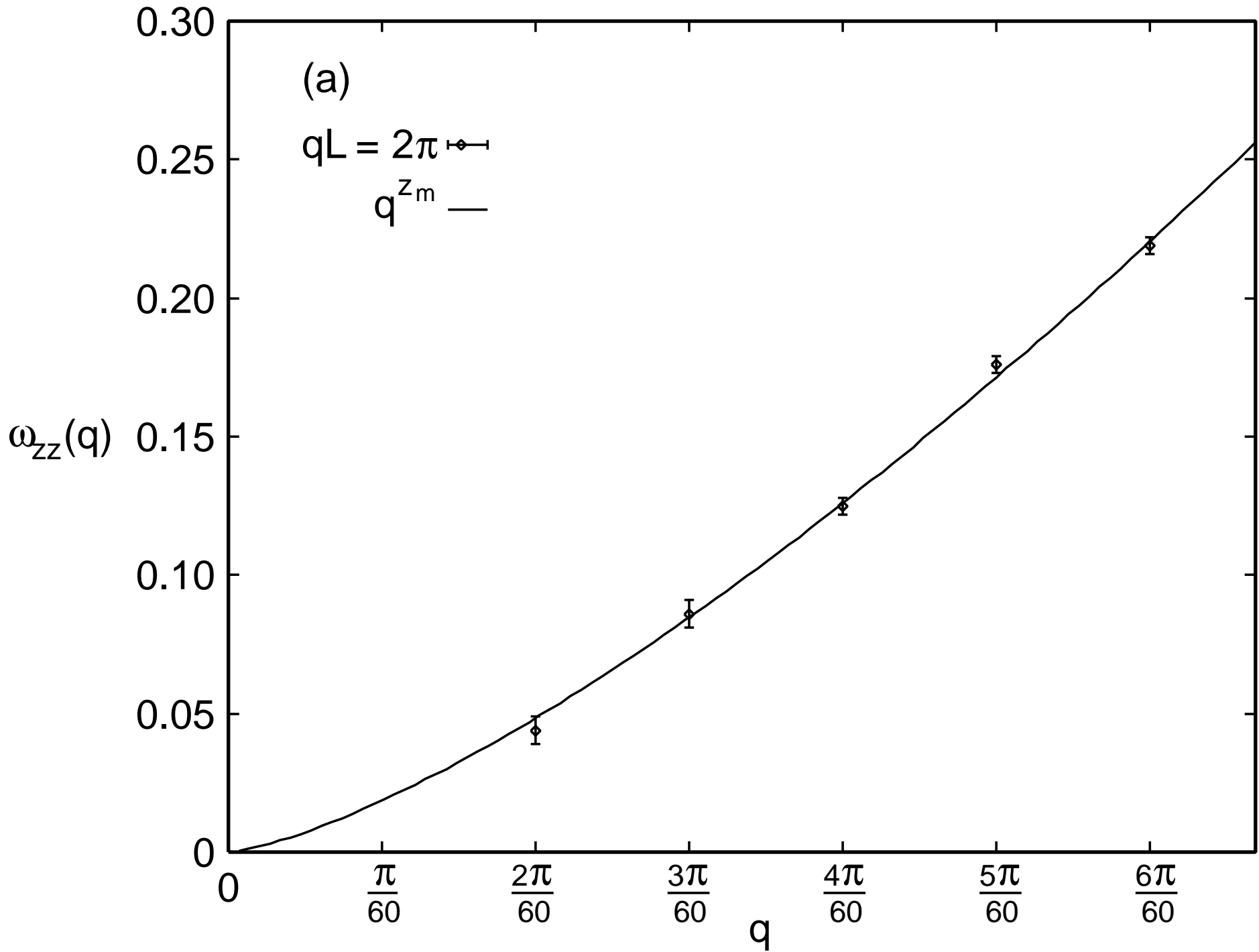
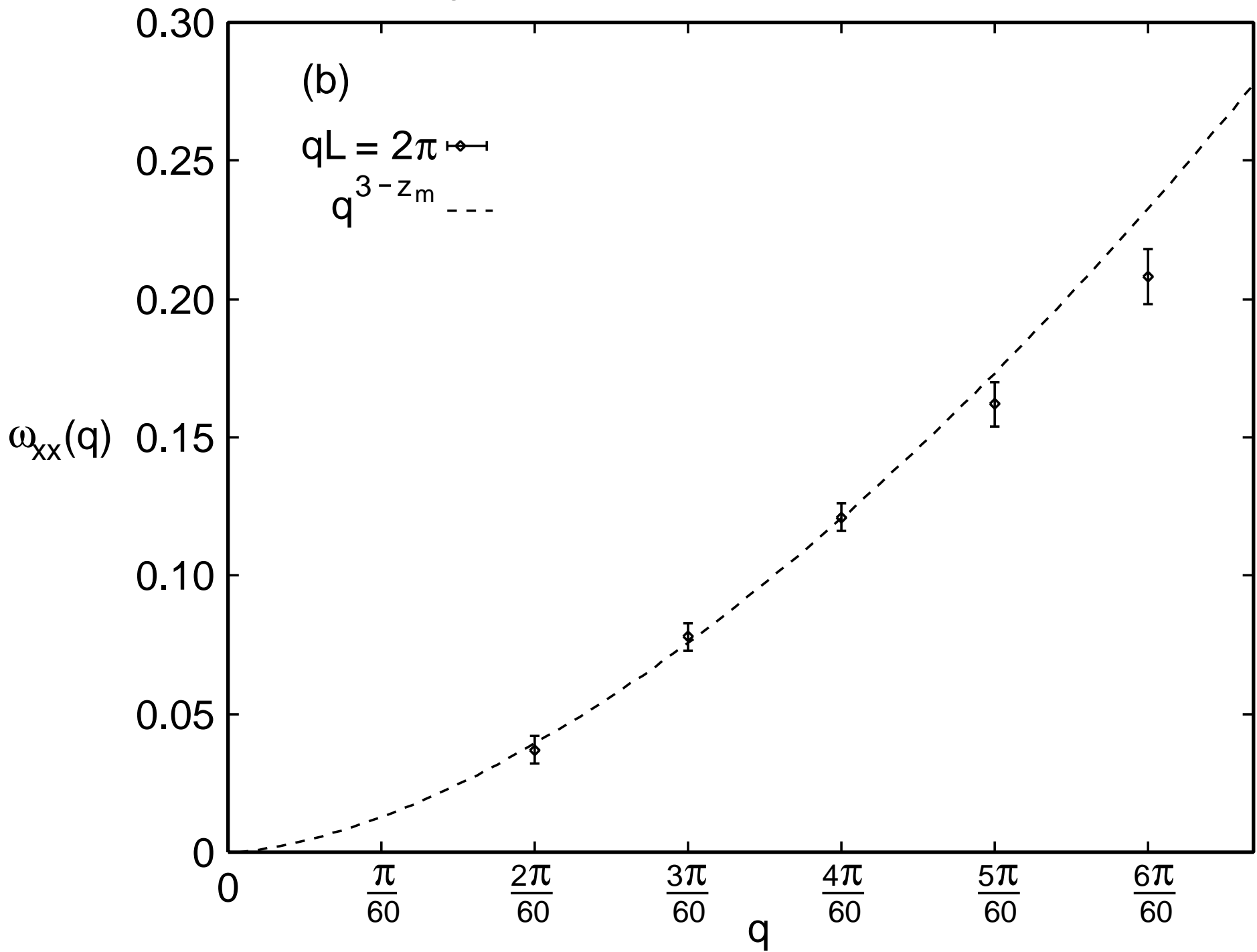
\begin{figure}[t]
\caption{Small $q$ behavior of the dispersion relations (a) $\omega_{zz}(q)$
and (b) $\omega_{xx}(q)$ in the $(100)$ direction for $T = T_c$. The solid
line is a power law fit to the data which yields $z_m = 1.38 \pm 0.05$. The
dashed line is a power law with the exponent $z_{\phi} = 3 - z_m = 1.62 \pm
0.05$. The data for the smallest systems $L = 20$ and $L = 24$ deviate
systematically from the power law.
\label{omzzTc}}
\end{figure}

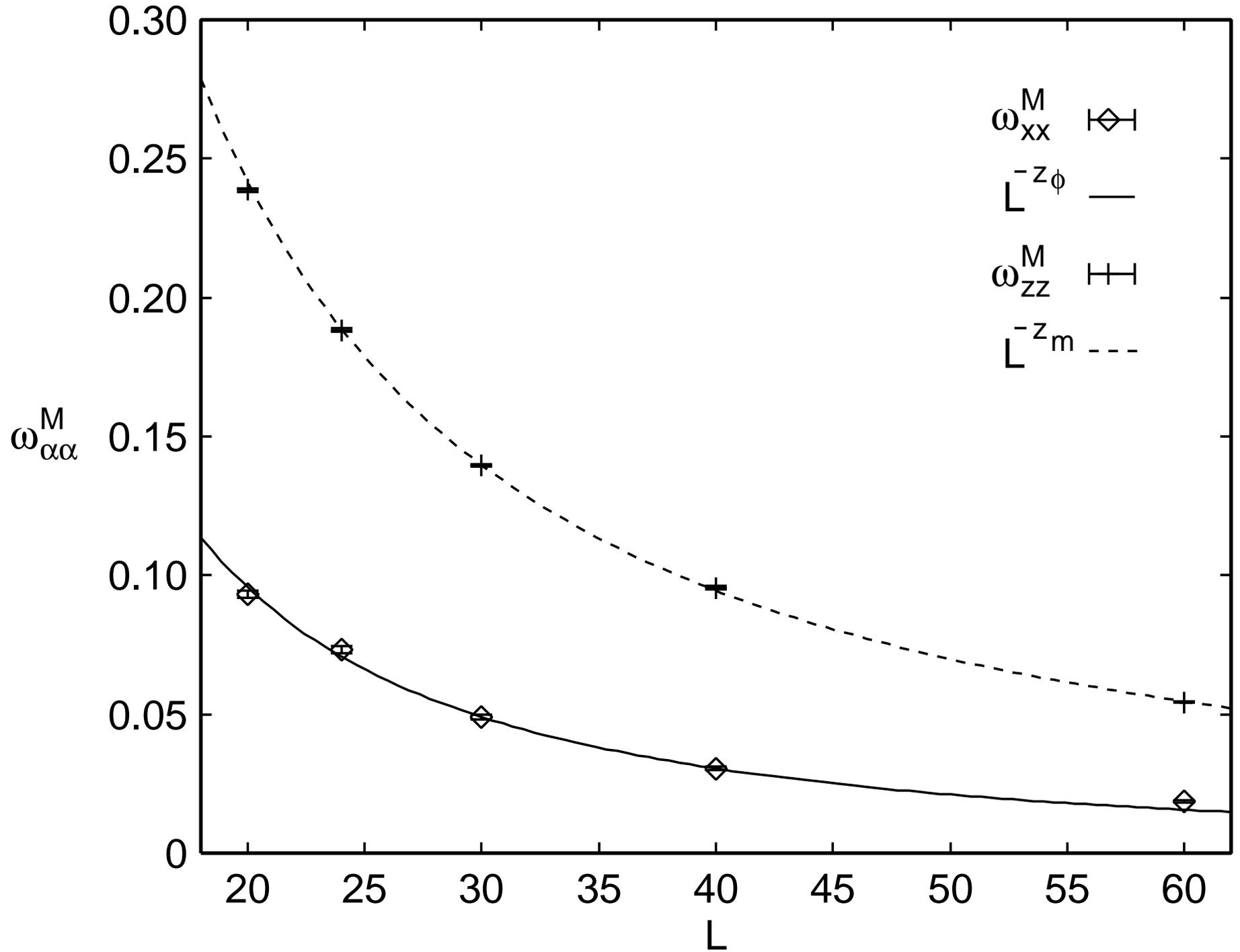
\begin{figure}[t]
\caption{Median frequencies $\omega_{xx}^M$ ($\Diamond$, solid line) and
$\omega_{zz}^M$ (+, dashed line) for $qL = 2\pi$. The solid and dashed lines
display fits to power laws (see Eq.(\protect\ref{ommscal})).
\label{ommed}}
\end{figure}

In order to provide numerical estimates of the dynamic exponents $z_m$
and $z_{\phi}$ first we analyze the dispersion relations $\omega_{zz}(q)$
and $\omega_{xx}(q)$ at the critical point as obtained from a Lorentz
fit to the spin wave peak of $S_{zz}(q,\omega)$ (see Eq.(\ref{Lzz}))
and a Lorentz fit to $S_{xx}(q,\omega)$ according to Eq.(\ref{Lxx}).
The line shape of $S_{xx}$ is dominated by a strong central peak (see
below) so that we restrict the analysis of the full dispersion relation
to $\omega_{zz}(q)$. The result in the $(100)$ direction is shown in
Fig.\ref{omTc} for $L = 24$ and $L = 40$. The data apparently collapse
onto a single curve and show a linear behavior for $q < \pi/2$ down to
$q \simeq \pi/20$, where $\omega(q)$ becomes nonlinear. From finite-size
scaling one expects the scaling form
\begin{equation}
\label{omscal}
\omega_{\alpha \alpha} = q^{z_{\alpha}} \Omega_{\alpha}(qL)
\end{equation}
at $T_c$ for sufficiently small q. For $qL = 2\pi$ and $L =$ 20, 24,
30, 40 and 60 we obtain an approximation of the small $q$ behavior
of $\omega_{zz}$ which is shown in Fig.\ref{omzzTc}(a). According to
Eq.(\ref{omscal}) $\omega_{zz}$ should vary as $q^{z_m}$ for fixed $qL$.
A least square fit to the data shown in Fig.\ref{omzzTc}(a) yields the
dynamic exponent
\begin{equation}
\label{zm}
z_m = 1.38 \pm 0.05
\end{equation}
which differs substantially from the mode-coupling prediction $z_m =
z_{\phi} = 1.5$ \cite{TFS91}. From the scaling relation $z_{\phi} = 3 - z_m$
\cite{DP78,Dohm78,Dohm91} we obtain $z_{\phi} = 1.62 \pm 0.05$ which can
be tested against the small $q$ behavior of $\omega_{xx}(q)$. The result
is shown in Fig.\ref{omzzTc}(b). For the larger systems $L = 30$, 40, and 60
the data agree with the power law, but for smaller lattice sizes $L = 20$
and $L = 24$ systematic deviations occur. If we exclude these smaller systems
from the power-law fit shown in Fig.\ref{omzzTc}(a) we find $z_m = 1.43
\pm 0.14$ which is at the upper error boundary of the previous estimate
given by Eq.(\ref{zm}). An alternative estimate of the dynamic exponents is
provided by the {\em median frequency} $\omega_{\alpha \alpha}^M$ which is
defined by the relation
\begin{equation}
\label{median}
\int_0^{\omega_{\alpha \alpha}^M} S_{\alpha \alpha}({\bf q},\omega)
d\omega = \textstyle{1 \over 2} G_{\alpha \alpha}({\bf q}).
\end{equation}
Note that we have normalized the dynamic structure factor such that
its integral over all frequencies yields the static structure factor.
We evaluate Eq.(\ref{median}) with the trapezoidal rule. Systematic
errors induced by this simple numerical integration method are smaller
than statistical uncertainties coming from the statistical errors of
$S_{xx}(q,\omega)$ and $S_{zz}(q,\omega)$. If the integration is performed
with Simpsons rule the same results are obtained within statistical
errors. According to Eq.(\ref{Sqwscal}) we expect the scaling behavior
\begin{equation}
\label{ommscal}
\omega_{\alpha \alpha}^M = L^{-z_\alpha} M_{\alpha \alpha}(qL)
\end{equation}
of the median frequency
at the critical point. Due to the presence of a strong central peak
in $S_{xx}(q,\omega)$ the median frequency $\omega_{xx}^M$ for $L = 60$
turns out to be of the same size as the frequency resolution of our
data. We therefore limit the analysis of $\omega_{xx}^m$ to the system
sizes $L =$ 20, 24, 30, and 40. The other median frequency $\omega_{zz}^m$
can be determined accurately for all system sizes. The result for $qL =
2\pi$ is displayed in Fig.\ref{ommed}. From a least square fit of
$\omega_{xx}^M$ to a power law for $L = 20$, 24, 30, and 40 we obtain
$z_\phi = 1.61 \pm 0.03$. If only the data for $L = 30$ and 40 are used
we obtain $z_\phi = 1.65 \pm 0.10$. Both estimates are consistent with
Eq.(\ref{zm}) and the scaling relation $z_\phi + z_m = 3$. The latter
estimate is displayed by the solid line in Fig.\ref{ommed}. The
corresponding fit of $\omega_{zz}^M$ for $L \geq 30$ yields $z_m = 1.35
\pm 0.01$ which is also within the error bar of our previous estimate
(see Eq.(\ref{zm}), dashed line in Fig.\ref{ommed}). The spread in the
values for the dynamic exponents obtained from the dispersion relations
for small $q$ and the median frequencies is considerable. In order to
reconcile all estimates with Eq.(\ref{zm}) almost the full width of the
error interval (one standard deviation) is needed. However, Eq.(\ref{zm})
represents a reasonable mean value of all estimates discussed so far and
we therefore adopt it as our final estimate for $z_m$ and use $z_\phi +
z_m = 3$ to determine $z_\phi$.

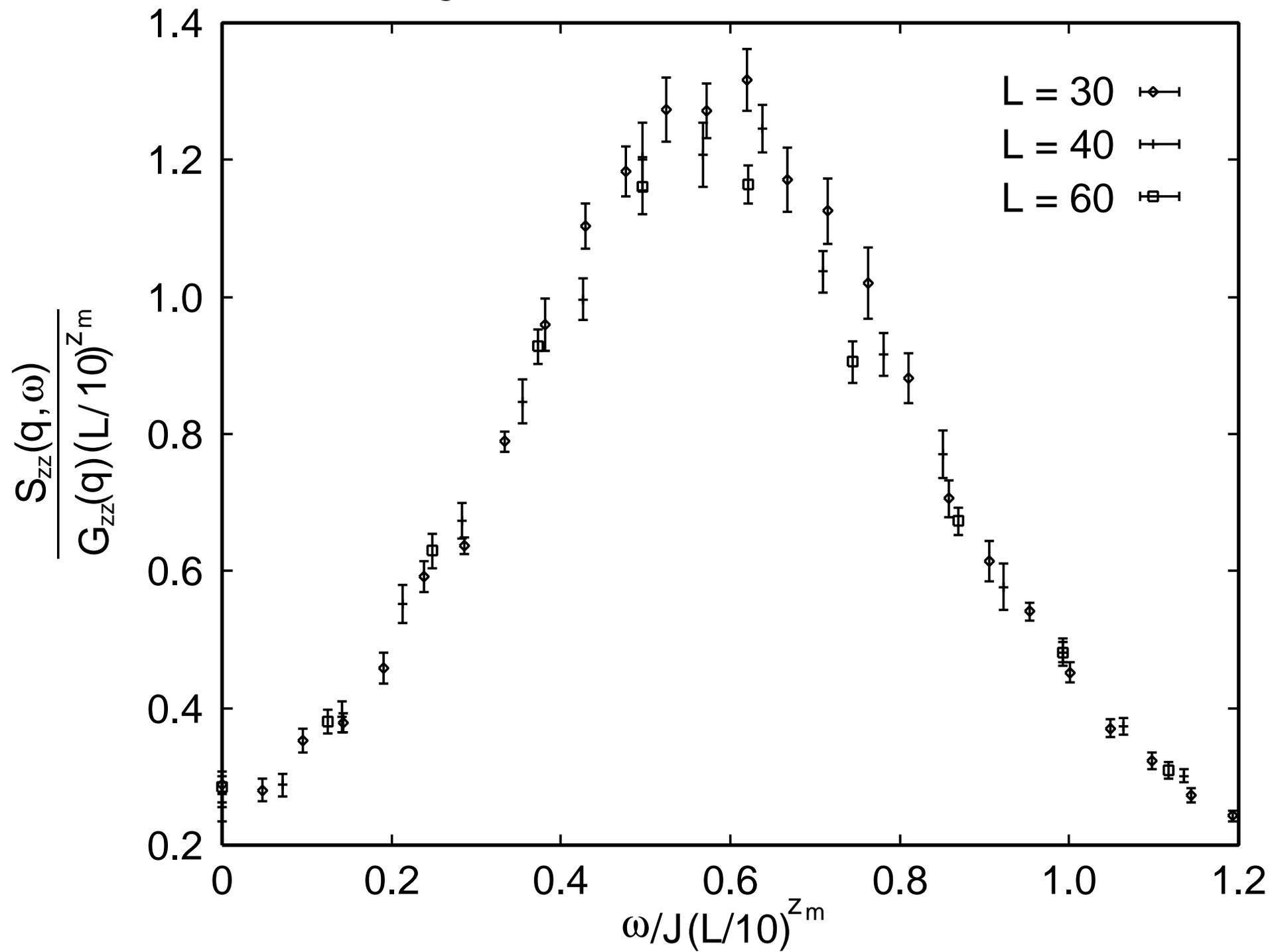
\begin{figure}[t]
\caption{Scaling plot of $S_{zz}(q,\omega)/\left((L/10)^{z_m} G_{zz}(q)
\right)$ versus $\omega/J (L/10)^{z_m}$ for $qL = 2\pi$ and system sizes
$L = 30$, 40, and 60.
\label{Szzscal}}
\end{figure}

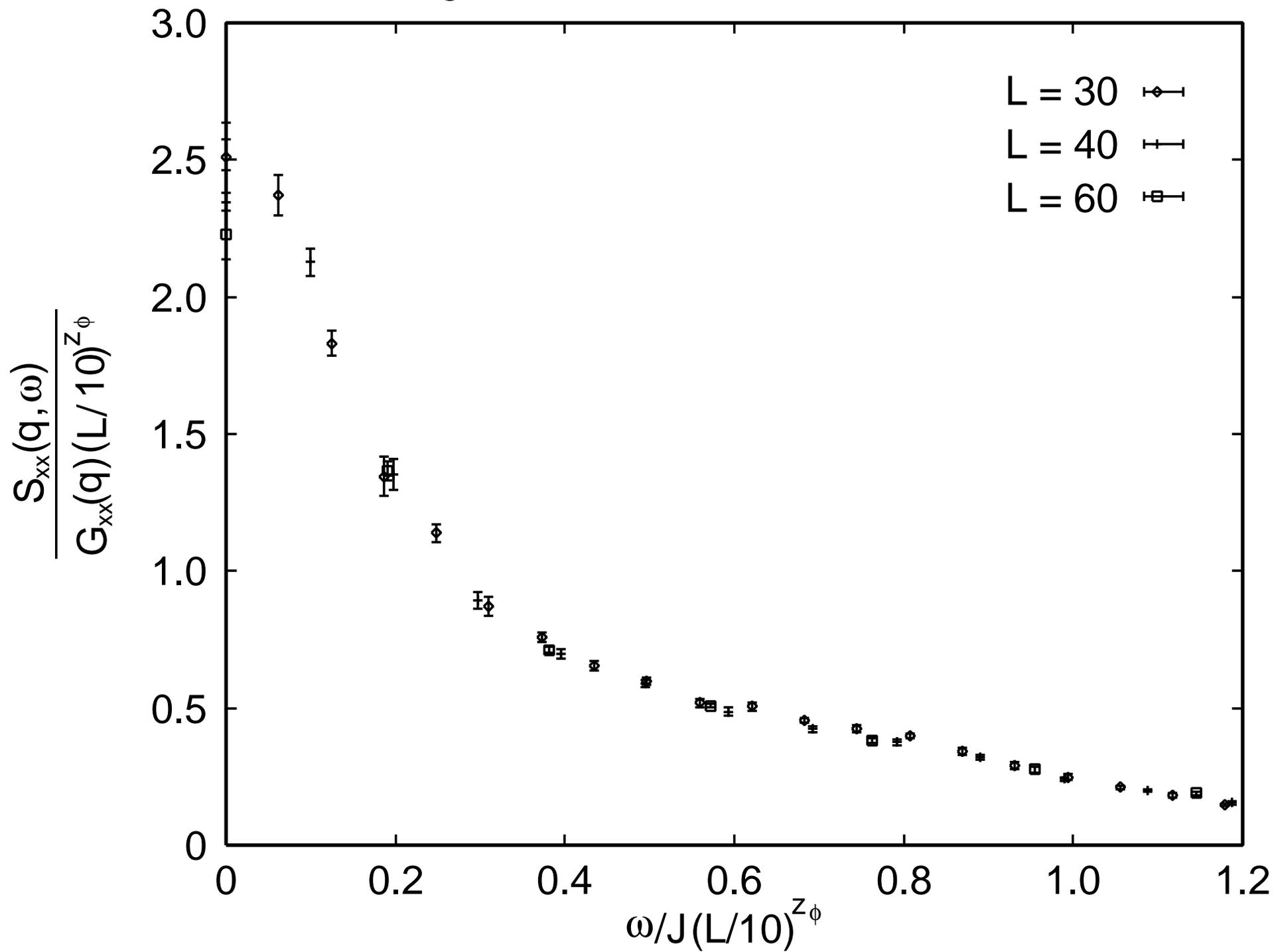
\begin{figure}[t]
\caption{Scaling plot of $S_{xx}(q,\omega)/\left((L/10)^{z_\phi}
G_{xx}(q)\right)$ versus $\omega/J (L/10)^{z_\phi}$ for $qL = 2\pi$ and
system sizes $L = 30$, 40, and 60.
\label{Sxxscal}}
\end{figure}

As a final test of Eq.(\ref{zm}) we use Eq.(\ref{Sqwscal}) in order to
obtain a scaling plot of the dynamic structure factor. For $qL = 2\pi$,
$L = 30$, 40 and 60 and $z_m = 1.38$ the resulting
scaling plot for $S_{zz}(q,\omega)$ is shown in Fig.\ref{Szzscal}. The
corresponding plot for $S_{xx}(q,\omega)$ with $z_\phi = 3 - z_m = 1.62$
is displayed in Fig.\ref{Sxxscal}. Data collapse can only be expected for
sufficiently small $\omega$. Scaling works quite well for all frequencies
up to the spin wave peak, where the intensity decreases slightly with
increasing $L$, but the error bars are still overlapping. The quality
of the data collapse in not so evident from $S_{xx}(q,\omega)$ (see
Fig.\ref{Sxxscal}). At $\omega = 0$ the scaled intensity for $L = 60$
deviates from the ones for $L = 30$ and 40 by about one standard deviation.
This could be an effect of the finite statistical sample. The line shape
of the central peak in $S_{xx}$ still fits rather well to a Lorentzian.
Only in the vicinity of $\omega = 0$ are the data also compatible with
the Gaussian lineshape expected from mode coupling theory \cite{TFS91}.

The above scaling analysis provides quite strong evidence for a violation
of dynamic scaling in the sense that two different dynamic exponents are
required in order to obtain scaling in the dispersion relations, the median
frequencies, and the two components of the dynamic structure factor.
However, we cannot prove from our data whether the estimate $\omega_w =
z_{\phi} - z_m = 0.24$, which indicates the violation of dynamic scaling,
corresponds to the transient exponent $\omega_w$ \cite{Dohm78,Dohm91} or
if the measured difference consitutes an {\em effective} exponent for system
sizes which are too small to ignore additional corrections. As pointed
out in Sec.I the presence of energy conservation in our spin dynamics
simulation gives rise to such corrections governed by the exponent
$\alpha / \nu$ \cite{HHS76}. On the basis of our data we cannot exclude the
possibility, that these corrections yield the dominant contribution to the
measured effective exponent $\omega_w$. For a clarification of this point
more information is needed from analytical theories of dynamic finite-size
scaling \cite{T98} and from simulations \cite{MK99}.

\section{Spin transport and thermal conductivity}
We have already mentioned in the introduction that the transport
properties of the XY model near the critical point provide lattice
analogues of the corresponding transport coefficients of $^4$He near
the $\lambda$-transition. The conserved out-of-plane component $M_z$
of the magnetization is the lattice analogue of the entropy density in
$^4$He and its associated transport coefficient therefore corresponds to
the thermal conductivity of $^4$He which is of experimental interest.
In principle the thermal conductivity can be extracted from an
extrapolation of the characteristic frequency $\omega_0$ of $S_{zz}$
given by $\omega_0^{-1} = {1 \over 2} S_{zz}(q,0)$ \cite{Dohm78,LHB74}
to $q = 0$ . However, in order to obtain a reliable extrapolation at
the critical point a very high momentum resolution for $q \to 0$ is
required and this can only be realized with very large systems (see
Fig.\ref{omTc} and Ref.\cite{LHB74}). We therefore resort to an
alternative approach already considered in Ref.\cite{LHB74},
we express the thermal conductivity by a current-current correlation
function for a suitably chosen current density ${\bf j}_k =
(j_{1,k},j_{2,k},j_{3,k})$ at lattice site $k$ \cite{KM63}.

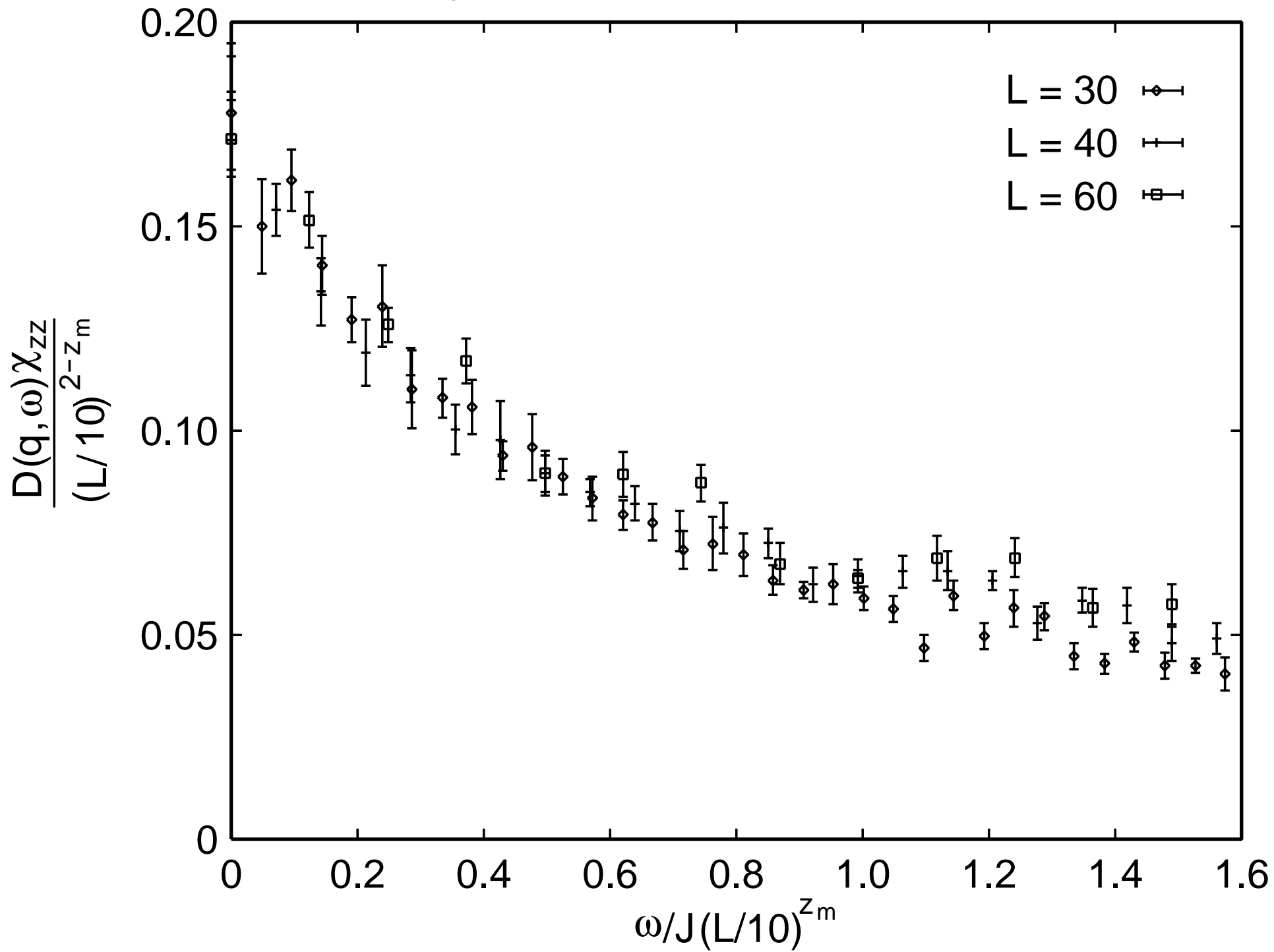
\begin{figure}[t]
\caption{Scaling plot of $D(q,\omega)\chi_{zz}/(L/10)^{2-z_m}$ versus
$\omega/J (L/10)^{z_m}$ for $q = 0$, $L = 30$, 40, and 60, and $z_m = 1.38$.
The value $D(q=0,\omega=0)$ corresponds to the thermal conductivity.
\label{D0wscal}}
\end{figure}

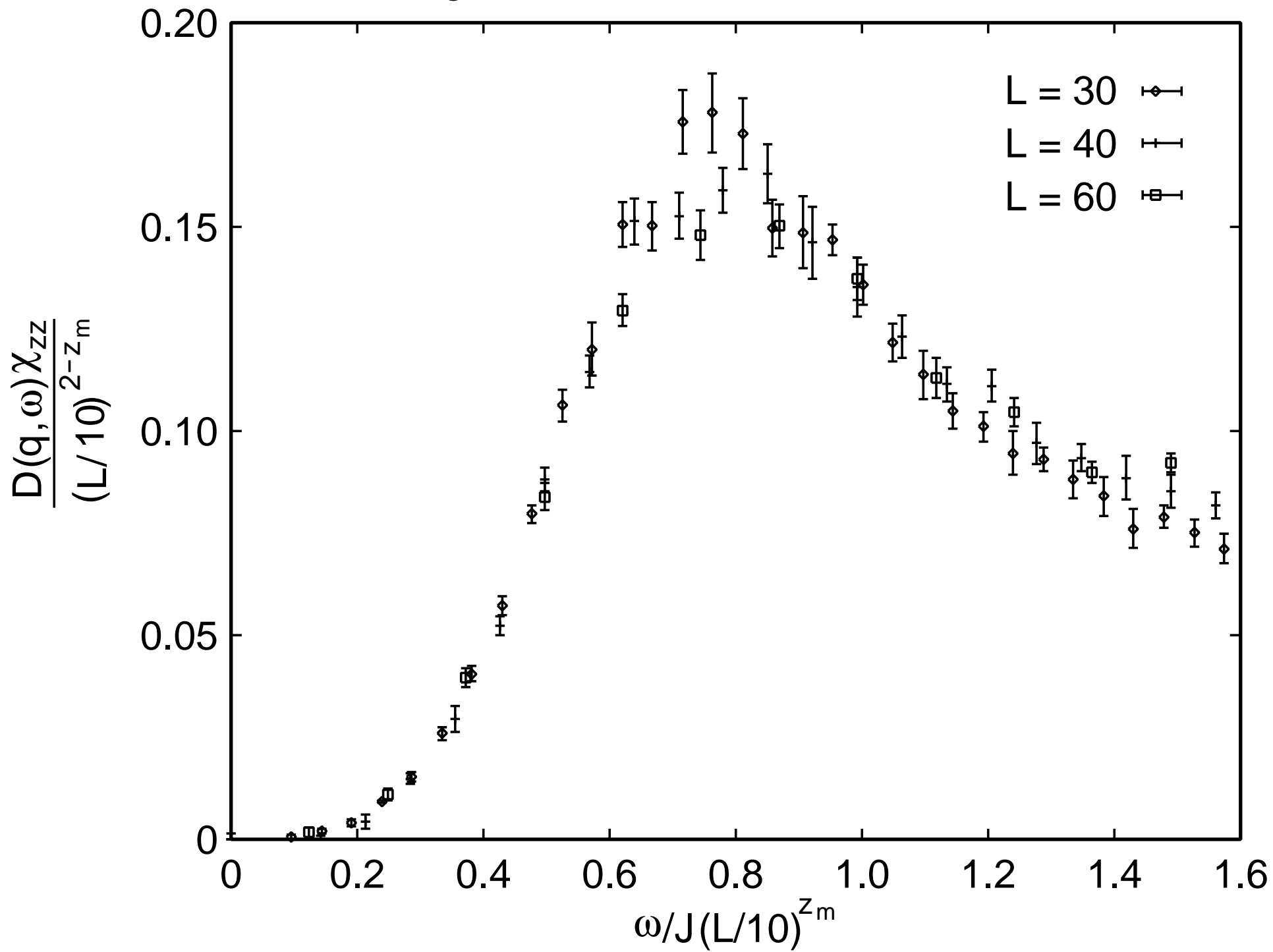
\begin{figure}[t]
\caption{Scaling plot of $D(q,\omega)\chi_{zz}/(L/10)^{2-z_m}$ versus
$\omega/J (L/10)^{z_m}$ for $qL = 2\pi$, $L = 30$, 40, and 60, and $z_m =
1.38$. $D(q,\omega)$ shows a strong spin wave resonance and it vanishes for
$\omega \to 0$.
\label{Dqwscal}}
\end{figure}

In order to identify the current density ${\bf j}_k$ we reexamine the
$z$-component of the equation of motion (see Eq.(\ref{XYeqmot})) which
reads
\begin{equation}
\label{XYeqmotz}
{d \over dt} S_k^z = -J \sum_{l \in NN(k)}
\left( S_l^x S_k^y - S_l^y S_k^x \right) ,
\end{equation}
where the sum is over all nearest neighbors of
lattice site $k$. We define the $i$th component $j_{i,k}$ $(i=1,2,3)$ of
the current density ${\bf j}_k$ associated with the lattice point $k$ by
\cite{LHB74}
\begin{equation}
\label{jik}
j_{i,k} \equiv J \left( S_k^y S_{k+e_i}^x - S_k^x S_{k+e_i}^y \right) ,
\end{equation}
where the notation $k + e_i$ denotes the nearest neighbor of the lattice
site $k$ in the $i$th lattice direction. For the case of the simple cubic
lattice studied here $e_1$, $e_2$, and $e_3$ can be visualized as the unit
vectors of a cartesian coordinate system aligned with the lattice. The
lattice divergence of the current density according to Eq.(\ref{jik}) at
lattice site $k$ is then given by
\begin{equation}
\label{divj}
\nabla \cdot {\bf j}_k = \sum_{i=1}^3 \left( j_{i,k} - j_{i,k-e_i} \right)
= J \sum_{l \in NN(k)} \left( S_l^x S_k^y - S_l^y S_k^x \right)
\end{equation}
which is just the negative r.h.s. of Eq.(\ref{XYeqmotz}). Note that the
lattice spacing has been set to unity. In the spirit of hydrodynamics
Eq.(\ref{XYeqmotz}) can be interpreted as second order discretization
of an equation of continuity of the form $\partial m_k /
\partial t = - \nabla \cdot {\bf j}_k$. The density $m_k \equiv S_k^z$ and the
current ${\bf j}_k$ defined by Eq.(\ref{jik}) are located on staggered meshes,
i.e., the current density component $j_{i,k}$ associated with lattice site
$k$ is located half way between $k$ and $k + e_i$, so that each of the finite
differences appearing in Eq.(\ref{divj}) is located at lattice site $k$
just as the l.h.s. of Eq.(\ref{XYeqmotz}). This displacement of spins and
currents on the lattice has no further consequences in our case, because
we have applied periodic boundary conditions in all lattice directions.
According to Eq.(44) of Ref.\cite{KM63} we find the expression
\begin{equation}
\label{Dqw}
D(q,\omega) = {1 \over k_B T \chi_{zz}}\ {2 \over \pi} \int_0^\infty \sum_i\
\langle j_{1,0}(0) j_{1,i}(t) \rangle \ \cos {\bf q} \cdot {\bf R}_i\
\cos \omega t\ dt
\end{equation}
for the transport coefficient $D(q,\omega)$, where ${\bf q} = (q,0,0)$ and
the same normalization as in Eq.(\ref{Sqw}) has been used. Note that the
out-of-plane static susceptibility $\chi_{zz} = \langle M_z^2 \rangle /
(k_B T L^3)$ needed for normalization in Eq.(\ref{Dqw})
is essentially constant as a function of system
size $L$ at fixed temperature. In the following we will analyze $D(q,\omega)$
only for $T = T_c$, where $\lambda \equiv D(0,0)$ corresponds to the thermal
conductivity measured in $^4$He experiments. According to Eq.(\ref{XYeqmotz})
the current density ${\bf j}_k$ has the scaling dimension $1 - z_m - d/2$,
because $\chi_{zz}$ has the scaling dimension zero. From
Eq.(\ref{Dqw}) we then find $2(1 - z_m - d/2) + d + z_m = 2 - z_m$ as the
scaling dimension of $D(q,\omega)$ so that naive finite-size scaling yields
$D(0,0) = \lambda \sim L^{2-z_m}$ at the critical point \cite{Dohm78}.
We therefore expect the scaling form
\begin{equation}
\label{DqwXzz}
D(q,\omega) k_B T_c \chi_{zz} = L^{2-z_m} \Delta(qL,\omega L^{z_m})
\end{equation}
at $T = T_c$, where $k_B T_c \chi_{zz}$ is just a normalization factor.
A corresponding scaling plot of $D(q,\omega) \chi_{zz}$ versus
$\omega/J L^{z_m}$ for $L = 30$, 40, and 60 and $q = 0$ is shown in
Fig.\ref{D0wscal}, where we have used our estimate $z_m = 1.38$.
The statistical noise in the data for $D(q,\omega)$ is
considerably larger than the noise in the data for the structure factor
(see also Ref.\cite{LHB74}). The spread of the data points in
Fig.\ref{D0wscal} is of the same magnitude as the scatter of the data in
each individual data set displayed. In view of these statistical
uncertainties our data scale reasonably well for small $\omega$ and confirm
the scaling exponent $2 - z_m \simeq 0.62$ (see Eq.(\ref{zm})) of the
transport coefficient $D(q,\omega)$. For $q \neq 0$ $D(q,\omega)$
scales accordingly as shown in Fig.\ref{Dqwscal} for $qL = 2\pi$. The position
of the spin wave maximum appears to be shifted to the right as compared to
the spin wave peak in $S_{zz}(q,\omega)$. According to Eq.(\ref{DqwXzz})
scaling is obtained up to a value of about 0.6 of the scaling argument for
both $q = 0$ and $qL = 2\pi$. According to Fig.\ref{Dqwscal} the spin wave
maximum itself appears to be outside the scaling regime for the system sizes
used here. For $q \neq 0$ $D(q,\omega)$ tends to zero as $\omega \to 0$
within the statistical errors.

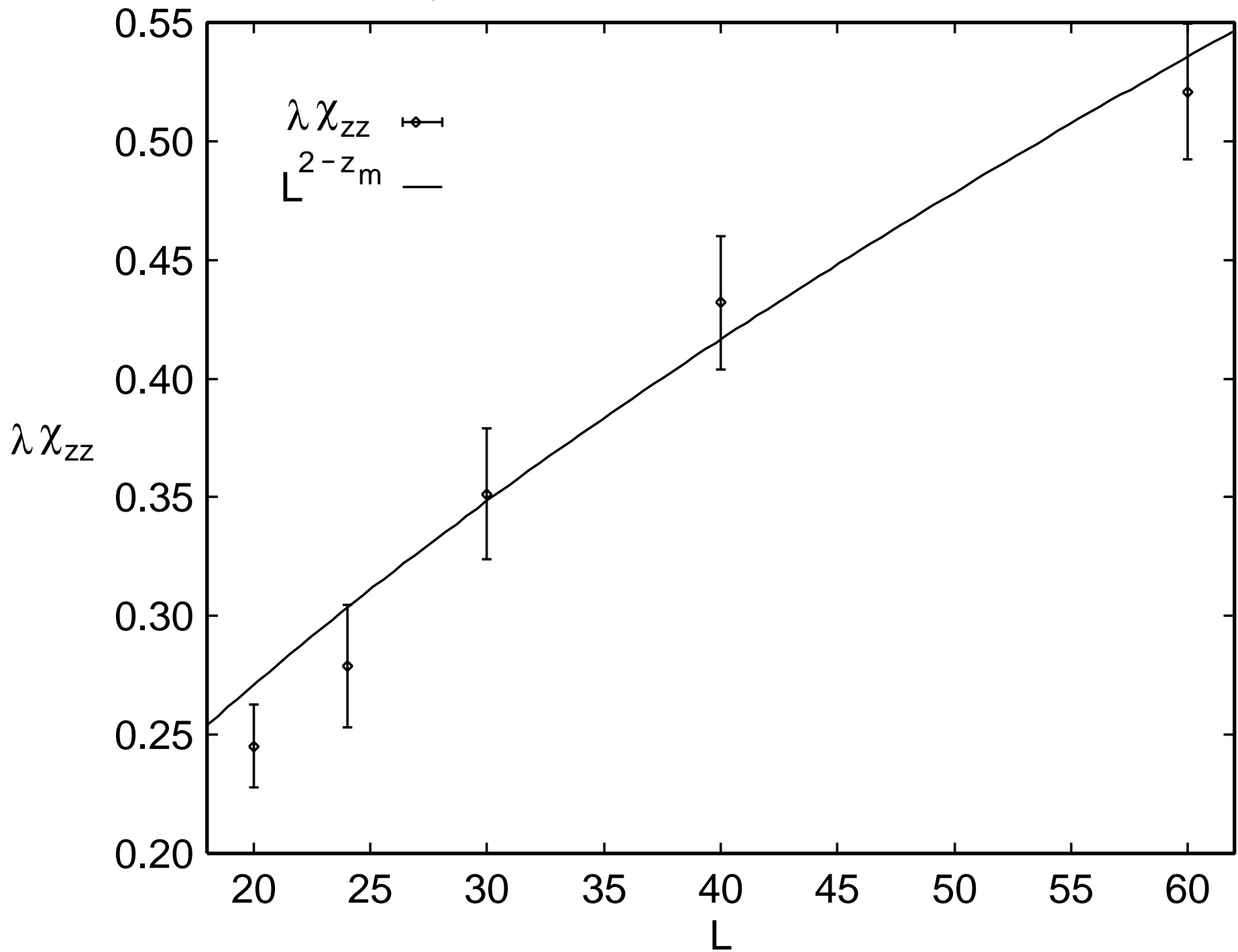
\begin{figure}[t]
\caption{Thermal conductivity $\lambda \chi_{zz}$ versus system size
$L$ for $L = 20$, 24, 30, 40, and 60. The solid line displays the
power law $L^{2-z_m}$ for $z_m = 1.38$ for comparison. The power law
represents the data reasonably well for $L \geq 30$.
\label{D00L}}
\end{figure}

The thermal conductivity $\lambda = D(0,0)$ is shown in Fig.\ref{D00L} as
a function of the system size. The overall behavior of $\lambda$
with $L$ is captured quite well by the expected power law (full line)
for $L \geq 30$ and even the data point for $L = 24$ is only one standard
deviation away. Systems with $L = 20$ or less may be too small to be in
the scaling regime. In view of the considerable statistical error in the
data slowly varying corrections to scaling as discussed in the previous
section cannot be identified. In any case more theoretical information in
needed in order to provide a reliable background for the interpretation of
spin dynamics data of transport coefficients in the critical regime
\cite{T98,MK99}.

\section{Summary and conclusions}
The easy-plane Heisenberg ferromagnet belongs to the XY universality
class which has been demonstrated by the evaluation of the critical
exponents and the static structure factor.
Unlike the standard plane rotator XY model the planar ferromagnet is
endowed with a reversible spin dynamics which can be efficiently simulated
by recently developed decomposition methods. Due to spatial and temporal
symmetries the spin dynamics of the planar ferromagnet is expected to
be in the same dynamic universality class as $^4$He near the superfluid
normal transition, but may have different corrections to scaling. The
data have been compared to field-theoretic and mode-coupling
predictions with the following main results.

\begin{enumerate}
\item Above the critical regime the spin dynamics data show a strong
Lorentzian central peak in the in-plane component $S_{xx}(q,\omega)$
in agreement with mode coupling theory. The out-of-plane component
$S_{zz}(q,\omega)$ displays a pseudo spin-wave peak which is also of Lorentzian
shape and becomes increasingly prominent as $q$ is increased. The presence
of this peak is in accordance with field-theoretic predictions for $T > T_c$.

\item Below the critical regime strong spin wave peaks occur in both
components of the dynamic structure factor. The shape of these peaks
is captured very well by Lorentians as expected from mode-coupling
theory. In addition to the spin wave peak $S_{xx}(q,\omega)$ displays
a central peak of Lorentzian shape and additional multi spin-wave peaks
which do not appear in $S_{zz}(q,\omega)$. For small $q$ the dispersion
relation $\omega(q)$ is linear in $q$ and the line widths $\Gamma_{xx}(q)$
and $\Gamma_{zz}(q)$ are quadratic in $q$ as expected from mode-coupling
theory. The qualitative agreement with the spin dynamics data for $d = 2$
suggests that the dynamics of the two-dimensional XY model may not be
captured by vortex dynamics theories.

\item At $T_c$, in contrast to mode-coupling theory and in agreement with
field-theoretic predictions, $S_{xx}(q.\omega)$ and $S_{zz}(q,\omega)$
require different dynamic exponents in order to obtain scaling: $z_m = 1.38
\pm 0.05$ and $z_\phi = 3 - z_m = 1.62 \pm 0.05$, whereas mode coupling
theory yields $z_m = z_\phi = 3/2$. The out-of-plane component
$S_{xx}(q,\omega)$ is dominated by a strong central peak. A shoulder at a
finite frequency indicates the presence of a spin wave signal. In
$S_{zz}(q,\omega)$ a strong spin wave peak remains the dominating feature.
The line shapes are still compatible with Lorentzians, the central peak in
$S_{xx}$ is only compatible with a Gaussian shape very close to $\omega = 0$.

\item The transport coefficient $D(q,\omega)$, which provides access to
the lattice analogue of the thermal conductivity of $^4$He within the
XY model, has been investigated at $T_c$. The statistical fluctuations of
the data are much stronger than those in the data for the structure factor
which makes the scaling analysis less unique. However, scaling in agreement
with the previously obtained dynamic exponents is obtained. The transport
coefficient also shows a strong spin wave resonance for finite $q$ and
vanishes for $\omega \to 0$ in this case. The thermal conductivity
$\lambda = D(0,0)$ scales with the system size in the expected way
within the error bars for sufficiently large systems.
\end{enumerate}

\acknowledgments
M. Krech gratefully acknowledges many helpful discussions with V. Dohm
and financial support of this work through the Heisenberg program of the
Deutsche Forschungsgemeinschaft. This research was supported in part by NSF
grant \#DMR - 9727714 and by the Computer Center of the RWTH Aachen.

\noindent

\begin{thebibliography}{99}
\bibitem{Amit78}
 D. J. Amit, {\em Field Theory, the Renormalization Group, and Critical
 Phenomena} (McGraw-Hill, New York, 1978).
\bibitem{Parisi88}
 G. Parisi, {\em Statistical Field Theory} (Addison-Wesley, Wokingham, 1988).
\bibitem{HH77}
 P. C. Hohenberg and B. I. Halperin, Rev. Mod. Phys. {\bf 49}, 435 (1977).
\bibitem{DPLMK98}
 D. P. Landau and M. Krech, J. Phys. Cond. Mat. {\bf 11}, R1 (1999).
\bibitem{Kawa76}
 K. Kawasaki in {\em Phase Transitions and Critical Phenomena} Vol. 5a,
 edited by C. Domb and M. S. Green (Academic, New York, 1976).
\bibitem{FS94}
 E. Frey and F. Schwabl, Adv. Phys. {\bf 43}, 577 (1994).
\bibitem{Keren94}
 A. Keren, Phys. Rev. Lett. {\bf 72}, 3254 (1994).
\bibitem{CL94}
 K. Chen and D.P. Landau, Phys. Rev. B {\bf 49}, 3266 (1994).
\bibitem{BCL96}
 A. Bunker, K. Chen, and D.P. Landau, Phys. Rev. B {\bf 54}, 9259 (1996).
\bibitem{RapLan96}
 D. C. Rapaport and D. P. Landau, Phys. Rev. E {\bf 53}, 4696 (1996).
\bibitem{HY79}
 K. Hirakawa and H. Yoshizawa, J. Phys. Soc. Jpn. {\bf 47}, 368 (1979).
\bibitem{HDJP86}
 M. T. Hutchins, P. Day, E. Janke, and R. Pynn, J. Mag. Mag. Mat. {\bf 54},
 673 (1986).
\bibitem{WZS94}
 D. G. Wiesler, H. Zabel, and S. M. Shapiro, Z. Phys. B {\bf 93}, 277 (1994).
\bibitem{KT72}
 J. M. Kosterlitz and D. J. Thouless, J. Phys. C {\bf 5}, L124 (1972);
 J. Phys. C {\bf 6}, 1181 (1973).
\bibitem{Huber82}
 D. L. Huber, Phys. Rev. B {\bf 26}, 3758 (1982).
\bibitem{Mertens}
 F. G. Mertens, A. R. Bishop, G. M. Wysin, and C. Kawabata, Phys. Rev. Lett.
 {\bf 59}, 117 (1987);
 F. G. Mertens, A. R. Bishop, M. E. Gouvea, and G. M. Wysin, J. de Physique
 C {\bf 8}, 1385 (1988);
 F. G. Mertens, A. R. Bishop, G. M. Wysin, and C. Kawabata, Phys. Rev. B
 {\bf 39}, 591 (1989).
\bibitem{Gouvea89}
 M. E. Gouvea, G. M. Wysin, A. R. Bishop, and F. G. Mertens, Phys. Rev. B
 {\bf 39}, 11840 (1989).
\bibitem{Volkel91}
 A. R. V\"olkel, F. G. Mertens, A. R. Bishop, and G. M. Wysin, Phys. Rev. B
 {\bf 43}, 5992 (1991).
\bibitem{DPLRG92}
 D. P. Landau and R. W. Gerling, J. Mag. Magn. Mater. 104 - 107, 246 (1992).
\bibitem{EvLan96}
 H. G. Evertz and D. P. Landau, Phys. Rev. B {\bf 54}, 12302 (1996).
\bibitem{Costa}
 J. E. R. Costa and B. V. Costa, Phys. Rev. B {\bf 54}, 994 (1996);
 B. V. Costa, J. E. R. Costa, and D.P. Landau, J. Appl. Phys. {\bf 81},
 5746 (1997).
\bibitem{TFS91}
 S. Thoma, E. Frey, and F. Schwabl, Phys. Rev. B {\bf 43}, 5831 (1991).
\bibitem{HHS76}
 B. I. Halperin, P. C. Hohenberg, and E. D. Siggia, Phys. Rev. B {\bf 13},
 1299 (1976).
\bibitem{DP78}
 C. De Dominicis and L. Peliti, Phys. Rev. B {\bf 18}, 353 (1978).
\bibitem{Dohm78}
 V. Dohm, Z. Phys. B {\bf 33}, 79 (1978);
 V. Dohm and R. A. Ferrel, Phys. Lett. {\bf 67A}, 387 (1978).
\bibitem{Ahlers68}
 G. Ahlers, Phys. Rev. Lett. {\bf 21}, 1159 (1968);
 W. Y. Tam and G. Ahlers, Phys. Rev. B {\bf 32}, 5932 (1985).
\bibitem{LSNCI96}
 J. A. Lipa, D. R. Swanson, J. A. Nissen, T. C. P. Chui, and U. E.
 Israelsson, Phys, Rev. Lett. {\bf 76}, 944 (1996).
\bibitem{Dohm91}
 V. Dohm, Phys. Rev. B {\bf 44}, 2697 (1991).
\bibitem{Wolff89}
 U. Wolff, Phys. Rev. Lett. {\bf 62}, 361 (1989).
\bibitem{Has90}
 M. Hasenbusch, Nucl. Phys. {\bf B333}, 581 (1990).
\bibitem{cluerr}
 A. M. Ferrenberg, D. P. Landau, and Y. J. Wong, Phys. Rev. Lett. {\bf 69},
 3382 (1992); L. N. Shchur and H. W. J. Bl\"ote, Phys. Rev. E {\bf 55},
 R4905 (1997).
\bibitem{AMFDPL}
 A. M. Ferrenberg and D. P. Landau, unpublished.
\bibitem{spindyn}
 M. Krech, A. Bunker, and D. P. Landau, Comp. Phys. Commun. {\bf 111}, 1
 (1998); J. Frank, W. Huang, and B. Leimkuhler, J. Comp. Phys. {\bf 133},
 160 (1997).
\bibitem{CFL93}
 K. Chen, A. M. Ferrenberg, and D. P. Landau, Phys. Rev. B {\bf 48}, 3249
 (1993).
\bibitem{GH93}
 A. P. Gottlob and M. Hasenbusch, Physica A {\bf 201}, 593 (1993).
\bibitem{LGZJ85}
 J. C. Le Guillou and J. Zinn-Justin, J. Phys. (France) Lett. {\bf 46},
 L137 (1985); R. Guida and J. Zinn-Justin, J. Phys. A {\bf 31}, 8103 (1998).
\bibitem{RF72}
 D. S. Ritchie and M. E. Fisher, Phys. Rev. B {\bf 5}, 2668 (1972).
\bibitem{T98}
 M. T\"opler, Diplomarbeit RWTH Aachen, 1998.
\bibitem{MK99}
 M. Krech, to appear in {\em Computer Simulation Studies in Condensed Matter
 Physics XII}, edited by D. P. Landau, S. P. Lewis, and H. B. Sch\"uttler
 (Springer Verlag, Heidelberg, Berlin, 1999).
\bibitem{LHB74}
 N. A. Lurie, D. L. Huber, and M. Blume, Phys. Rev. B {\bf 9}, 2171 (1974).
\bibitem{KM63}
 L. P. Kadanoff and P. C. Martin, Ann. Phys. {\bf 24}, 419 (1963).
\end{thebibliography}
\end{document}